\documentclass[useAMS,usenatbib,usegraphicx]{mn2e}
\usepackage{times}
\usepackage{natbib}
\usepackage{journals}
\usepackage{xcolor}
\usepackage{amsmath}

\begin{document}
\bibliographystyle{mn2e}

\title[ALMA detection of C\,{\sc II} at $z=4.6$]{ALMA detection of a disc-dominated [C\,{\sc II}] emission line at $z=4.6$ in the luminous QSO J1554+1937}
\author[Kimball et al.]{Amy E. Kimball,$^1$\thanks{E-mail: Amy.Kimball@csiro.au} Mark Lacy,$^2$ Carol J. Lonsdale$^2$ and J.-P. Macquart$^3$\\
$^1$CSIRO Astronomy and Space Science, Australia Telescope National Facility, PO Box 76, Epping, NSW 1710, Australia\\
$^2$National Radio Astronomy Observatory, 520 Edgemont Rd., Charlottesville, VA 22903, USA\\
$^3$ICRAR/Curtin University GPO Box U1987, Perth, WA 6845, Australia}

\maketitle

\begin{abstract}
We present observations and analysis of an unusual [C\,{\sc II}] emission line in the very luminous QSO SDSS J155426.16+193703.0 at $z\sim4.6$. The line is extremely broad (FWHM 735~km~s$^{-1}$) and seems to have a flat-topped or double-peaked line profile. A velocity map of the line shows a gradient across the source that indicates large-scale rotation of star-forming gas. Together, the velocity map and line profile suggest the presence of a massive rotating disc with a dynamical mass $M_\mathrm{dyn}\ga5\times10^{10}~M_\odot$. Using the assumption of a rotating disc origin, we employ an empirical relation between galaxy disc circular velocity and bulge velocity dispersion ($\sigma$) to estimate that $\sigma>310$~km~s$^{-1}$, subject to a correction for the unknown disc inclination. This result implies that this source is consistent with the local $M$--$\sigma$ relation, or offset at most by an order of magnitude in black hole mass. In contrast, the assumption of a bulge origin for the [C\,{\sc II}] emission line would lead to a conclusion that the black hole is nearly two orders of magnitude more massive than predicted by the $M$--$\sigma$ relation, similar to previous findings for other high-redshift QSOs. As disc rotation may be a common origin for [C\,{\sc II}] emission at high redshifts, these results stress that careful consideration of dynamical origins is required when using observations of this line to derive properties of high-redshift galaxies.

\end{abstract}


\section{Introduction}
\label{sec:intro}

With the recent inauguration of the Atacama Large Millimeter/Sub-millimeter Array (ALMA) telescope, the astronomical literature has shown a surge in studies of the 157.74~$\mu$m forbidden $^2\mathrm{P}_{3/2} \rightarrow~^2\mathrm{P}_{1/2}$ fine structure line ([C\,{\sc II}]) of ionized carbon. This line arises from photo-dissociation regions in the cold neutral interstellar medium (ISM), marking it as a tracer of cool gas and star formation \citep[for a recent review, see][]{carilli_review13}. Furthermore, [C\,{\sc II}] is typically the brightest emission line in the ISM: in ultra-luminous infrared galaxies (ULIRGs), this single line can account for $\sim0.1$--1 per cent of the entire far-infrared (FIR) luminosity \citep{luhman98}. Therefore, [C\,{\sc II}] serves as an incredibly powerful tool for probing the dynamics of star formation (SF) activity in galaxies. This statement is especially true at high redshifts owing to the high luminosity of [C\,{\sc II}] compared to other galaxy emission lines, as well as to the high angular-resolution of ALMA.

The [C\,{\sc II}] line will be particularly valuable for observing quasi-stellar objects (QSOs, also commonly referred to as quasars), because it is one of few tools to directly observe the host galaxies of these luminous sources. QSOs are the signposts of actively accreting super-massive black holes (SMBHs), and at most frequencies they vastly outshine their host galaxies. 
The SMBH's accretion disc radiates strongly in the optical and ultraviolet (UV) regimes; the disc's corona up-scatters disc photons into the X-ray regime; and hot dust close to the SMBH radiates thermally with a peak in the mid-infrared (MIR). However, in the FIR and sub-millimetre regimes, cool dust heated by star formation processes in the host galaxy has a substantial contribution to the emission rate. Therefore, sub-millimetre emission lines such as [C\,{\sc II}] are an excellent way to observe the host galaxy in a frequency regime that may be less dominated by the SMBH.

Using [C\,{\sc II}] to investigate high-redshift sources is a recently established field of study. \citet{maiolino05} were the first to obtain a [C\,{\sc II}] detection in a high-redshift ($z=6.42$) QSO, which they achieved using the IRAM 30-m single-dish telescope. Many early resolved detections of high-redshift [C\,{\sc II}] were obtained with the IRAM Plateau de Bure Interferometer \citep[PdBI;][]{walter09,maiolino12,gallerani12,venemans12,cicone15}. These observations have indicated that [C\,{\sc II}] in high-redshift QSO host galaxies is found within 1~kpc of the galaxy nucleus. The presence of [C\,{\sc II}] on small spatial scales suggests that vigorous star formation is taking place in the central regions of these host galaxies, although a quantitative estimate of the overall star formation rate (SFR) based only on [C\,{\sc II}] luminosity would be unreliable \citep{ds4mnras}. IRAM observations of [C\,{\sc II}] revealed typical velocity widths of up to several hundred kilometres per second. However, in observations of QSO J1148+5251, \citet{maiolino12} and \citet{cicone15} detected a second, much broader component  of the line (FWHM~$\sim2000$~km~s$^{-1}$) extending tens of kiloparsecs from the nucleus, suggesting the presence of a rapid, massive gas outflow powered by the QSO.

ALMA is now poised to revolutionize [C\,{\sc II}] studies of QSO host galaxies, as its incredible sensitivity allows high signal-to-noise detections with integration times of just a few minutes. In ALMA {\it Early Science} mode, \citet{willott13} detected [C\,{\sc II}] at $z=6.4$ in QSO CFHQS~J0210$-$0456, obtaining similar results to the earlier IRAM observations. Willott~et al.~also obtained a new result: they observed a smooth velocity gradient of 100~km~s$^{-1}$ across a spatial scale of 6~kpc, which could potentially originate from rotation of a galaxy-wide disc. \citet{wang13} followed with a larger study in ALMA Cycle 0, detecting the [C\,{\sc II}] line in five other $z\sim6$ QSO host galaxies. They also observed a velocity gradient in four of their five targets, with line luminosities in the range 2--8$\times10^9~\mathrm{L_\odot}$ and FWHM in the range 200--500~km~s$^{-1}$.

These emerging high-redshift QSO studies will be crucial for probing the role that QSOs play in galaxy evolution. A correlation between the mass of the SMBH and the bulge stellar velocity dispersion, known as the $M$--$\sigma$~relation \citep[e.g.,][]{kormendyRichstone95,gultekinRelations09}, suggests a connection between the growth of a galaxy's bulge and its SMBH, despite the fact that the radius of influence of a SMBH is significantly smaller than the bulge radius \citep{ferrareseMerritt00}. It is a fairly tight relation, but the amount and origin of the scatter remain controversial \citep[e.g.,][]{graham09,graham11,graham15}. The most commonly proposed theory for this correlation is the presence of a `feedback' mechanism, wherein the SMBH exerts radiative or mechanical energy, in proportion to its accretion rate, that slows or halts star formation in the host galaxy. The two most widely supported theories for the feedback mechanism are (a) radiatively driven gas outflows in luminous sources and (b) jet momentum in radio-loud sources. An alternative theory to QSO feedback is that the $M$--$\sigma$ relation arises naturally from the statistics of mass averaging in galaxy mergers \citep{peng07,jahnkeMaccio11}. As presented in \citet{kormendyHo_review}, the local $M$--$\sigma$ relation is established from observations of about 50 nearby galaxies whose SMBH mass can be reliably determined through direct dynamical measurements. Extension to high redshifts is tricky, as the relevant parameters must be determined through secondary methods, and often involves an extrapolation that was originally calibrated on the local relation. Where such high-redshift studies have been performed, they indicate that the ratio of SMBH mass to bulge properties is higher than the local ratio \citep[e.g.,][]{peng06,wang10,willott13}. Such results have been interpreted as indication that SMBHs have an early, rapid growth phase that precedes galaxy growth \citep[e.g.,][]{mclure06,reines11nature,willott13}.

Studying the most luminous QSOs is another approach to these topics. Naturally, selection biases necessitate that high-redshift samples are intrinsically quite luminous; however, extremely luminous QSOs are observed across a range of redshifts, although they are too rare to manifest at $z\la1$. The most luminous QSOs likely represent the most massive SMBHs undergoing strong growth, and/or less luminous SMBHs experiencing super-Eddington accretion rates. While there is significant controversy as to whether QSOs are truly associated with mergers \citep[e.g.,][and references therein]{treister10,CANDELS12,heckmanBest14}, there is strong evidence that mergers do trigger the most luminous ones \citep{treisterEtal12}. Additionally, there is evidence that strong gas outflows are ubiquitous, or nearly so, in the host galaxies of luminous QSOs \citep[e.g.,][]{green06,krongold10,liuZakamskaGreene14}. Such outflows complicate these studies, as lines that indicate outflows are often the same emission lines that are used to estimate SMBH mass (e.g., C\,{\sc IV}) or galaxy dynamics (e.g., [C\,{\sc II}]) at high redshifts \citep{richardsEtal11,maiolino12,cicone15}. The most luminous sources are rare objects, pushing the boundaries of our theories and understanding, and thus are very interesting to study but also difficult to interpret.

In this manuscript, we present an ALMA Cycle 1 detection of an extremely broad [C\,{\sc II}] line in an extremely luminous ($L_\mathrm{bol}>10^{14}~\mathrm{L_\odot}$) QSO. (Results and implications from the continuum observations of our full QSO sample will be presented in a later publication.) Section~\ref{sec:observations} of this manuscript presents the target selection and observations. Section~\ref{sec:data} presents the data reduction and analysis. In Section~\ref{sec:discussion}, we present interpretations of the line, provide estimates for the bulge and disc properties, and discuss this source in the context of the $M$--$\sigma$ relation. We conclude in Section~\ref{sec:summary}. 

Throughout the remainder of this paper, we use a flat $\Lambda$ cosmology where $\Omega_\Lambda=0.7$ and $H_0=70$~km~s$^{-1}$~Mpc$^{-1}$.

\section{Target selection and ALMA observations}
\label{sec:observations}

We first identified luminous QSOs as potential targets from the multi-wavelength QSO catalog of \citet{krawczyk13}, which was compiled mainly from the Sloan Digital Sky Survey (SDSS) Seventh Data Release Quasar Catalog \citep{dr7quasars} and various broad-band sky surveys ranging from MIR to X-ray. Following the recommendation of Krawczyk et al., we define a QSO's bolometric luminosity as the energy in the (rest-frame) wavelength range 1~$\mu$m--0.62~nm (2~keV). This range, which is dominated by the accretion disc luminosity, corresponds to the intrinsic energy of the QSO prior to re-processing of any kind, and therefore gives the best estimate of bolometric luminosity when there is no significant covering fraction or absorbing material along the line of sight (as for Type~1 QSOs).
Of $\sim$120,000 optically identified QSOs in the Krawczyk et al.~catalog, fewer than 0.1 per cent have bolometric luminosity greater than $10^{14}\mathrm{L_\odot}$, meaning that they are as or more luminous than the most luminous known obscured QSOs \citep[][C.~Lonsdale et al., in preparation]{eis4mnras, wu12_HLIRG, bridge13}. With a redshift of $z=4.612$, the QSO SDSS J155426.16+193703.0 (hereafter J1554+1937) has the highest redshift of these most luminous QSOs from the Krawczyk et al.~catalog.

\begin{figure*}
\includegraphics[width=7in]{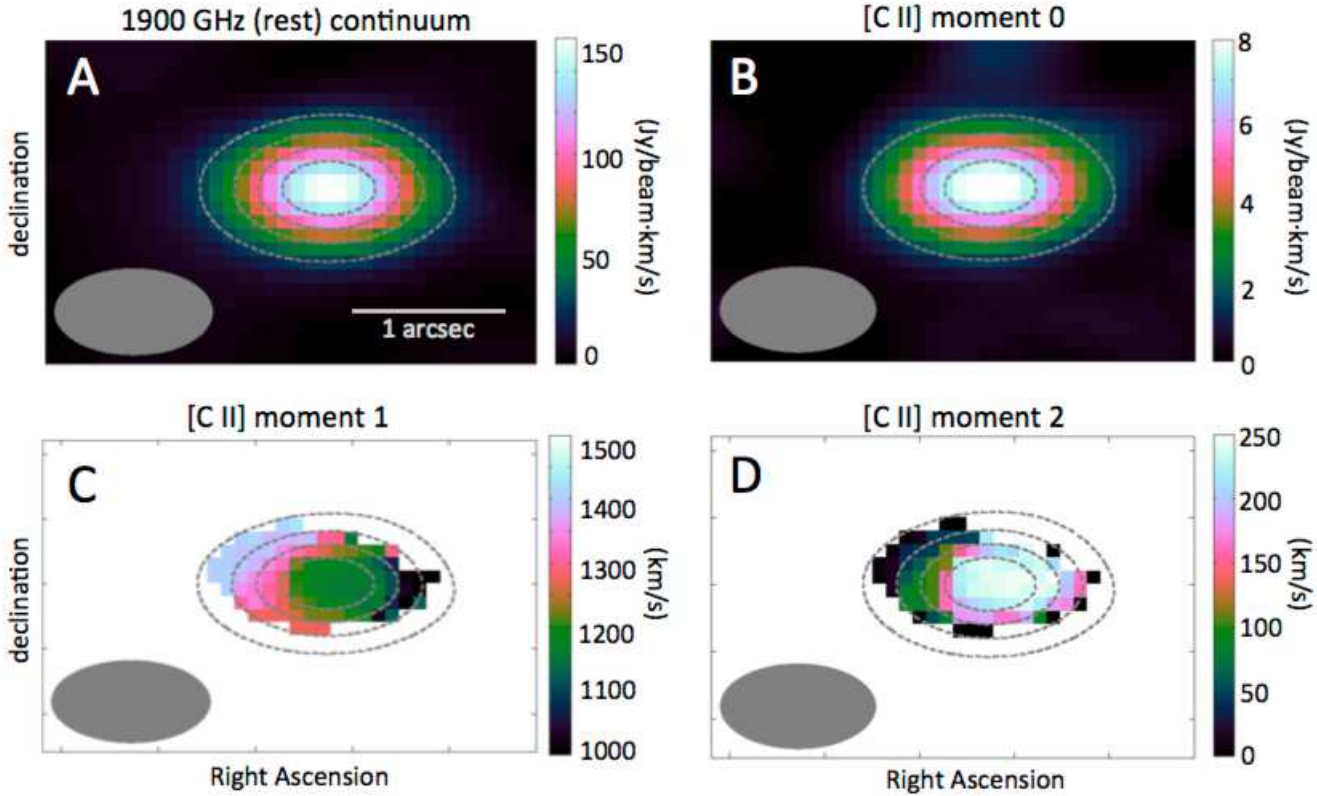}
\caption{
We show 1900~GHz (158~$\mu$m) continuum (Panel~A) and moment maps (as labelled; Panels~B through~D) of the [C\,{\sc II}] line of J1554+1937. The grey dashed contours show the continuum profile in each panel (the color-scale and contours in Panel~A are equivalent). The filled grey ellipse indicates the shape and orientation of the synthesized beam (analogous to `point-spread function' in optical/IR observations). The emission shown in Panels~C and~D is restricted to regions where the signal-to-noise ratio in the [C\,{\sc II}] data cube is greater than~4.}
\label{fig:maps}
\end{figure*}

J1554+1937 has a bolometric luminosity $L_\mathrm{bol}=1.3\times10^{14} \mathrm{L_\odot}$. There is no known evidence to suggest that J1554+1937 is lensed (and thus has an artificially brightened apparent luminosity). Other than its extremely high luminosity it does not seem to be an unusual QSO. Its broad-band spectral energy distribution (SED) and optical spectrum are typical for a high-redshift QSO. It is undetected in the NRAO--VLA Sky Survey \citep{nvss} and in the Faint Images of the Radio Sky at 20 cm (FIRST) survey \citep{first}; thus its 1.4-GHz flux density is less than $\sim1$~mJy. The typical criterion to classify a QSO as radio-loud or radio-quiet is the radio-to-optical ratio, $R$, originally defined in the observed frame: radio-loud QSOs generally have $R$ values in the range 10--1000, while radio-quiet QSOs have $R<1$ \citep{kellermann89}. J1554+1937 has an observed-frame radio-to-optical ratio $R\la40$ (determined from the FIRST survey upper limit and the $g$-band magnitude), meaning that it may be radio-quiet, radio-intermediate, or at the low end of the radio-loud range.

The redshift of J1554+1937 makes its [C\,{\sc II}] line accessible to ALMA's Band 7 (275--373~GHz). QSO emission lines are quite broad \citep[typically $>1000$~km~s$^{-1}$ even for the narrow lines; e.g.,][]{hao05}, and therefore do not require high spectral resolution. We placed the centres of two of the 2-GHz spectral windows at 337 and 339 GHz, covering the full 4-GHz-wide lower sideband with a spectral resolution of $\sim30$~km~s$^{-1}$. A small frequency range (19.75~MHz) at 338~GHz is unusable due to low sensitivity at the spectral window edges. Two other spectral windows covered the frequency range 348--352~GHz.

In interferometric observations, there is a necessary trade-off between better angular resolution (with a more extended array configuration) and better sensitivity (with a more compact array configuration). The primary motivation for our observations was a continuum detection experiment (the results of which will be discussed in a forthcoming paper), and for this reason we opted for an array configuration that would achieve high sensitivity at the cost of spatial resolution. The observations were performed using a compact configuration, resulting in angular resolution of approximately 1~arcsec at 340~GHz.

J1554+1937 was observed for 5.2 minutes with 25 available antennas. Titan was used as the absolute flux calibration source and nearby QSO J1619+2247 (6.6 degrees distant) as the phase calibrator.


\section{Data Analysis}
\label{sec:data}

Initial data reduction was performed by staff at the North American ALMA Science Center\footnote{\tt https://science.nrao.edu/facilities/alma} (NAASC). The data package provided to us by the NAASC included Python scripts that would execute the basic data reduction and calibration; the data package also included preliminary images. The scripts were executed in version 4.2 of {\sc CASA}\footnote{{\tt http://casa.nrao.edu/}} \citep{casa}. We performed minor modifications to the scripts in order to reduce the number of frequency channels flagged at the spectral window edges. We produced our own set of de-convolved images using {\sc CASA}'s {\tt clean} task, for which we used Briggs weighting with parameter setting {\tt robust=0.5}. We performed a single round of phase-only self-calibration using solutions from the combined continuum and line data. The synthesized beam size (analogous to the `point-spread function' in optical/IR observations) is 1\farcs2 $\times$ 0\farcs7, corresponding to $7.8\times4.6$~kpc at $z=4.6$.

The continuum map (in the frequency ranges 336--337~GHz, 339--340~GHz, and 348--352~GHz) is shown in Panel~A of Fig.~\ref{fig:maps}; the rms noise level in the continuum image is approximately 0.15~mJy~beam$^{-1}$. To estimate the source size, we performed a Gaussian fit to the data in the UV-plane using the {\sc CASA} task {\tt uvmodelfit}, which yielded a de-convolved source size of $310\pm30\times260\pm70$~mas ($\sim2.0\times1.7$~kpc) indicating that the source is marginally resolved spatially. Based on the maximum baseline length of $\approx400$~m, the smallest angular scale we could resolve at 338~GHz is 0\farcs46.
The integrated flux density resulting from the fit is $12.70\pm0.12$~mJy at $\sim344$~GHz. At a redshift of $z=4.6$, that flux density corresponds to a luminosity of $\sim2.3\times10^{12}~\mathrm{L}_\odot$ at a rest frequency of 1930~GHz (156~$\mu$m). This value is three orders of magnitude more luminous than the upper limit of the 1.4-GHz luminosity (see Sec.~\ref{sec:observations}); therefore it is clear that the 1930-GHz continuum emission is due to dust rather than a radio jet.

We isolated the line emission by subtracting the continuum from the data cube. An intensity map (moment~0 map) of the line using the frequency range 337.3--338.2~GHz is shown in Panel~B of Fig.~\ref{fig:maps}; the rms noise level in this image is approximately 0.4~Jy~beam$^{-1}$~km~s$^{-1}$. We again used {\sc CASA}'s {\tt uvmodelfit} task to estimate the spatial extent of the line-emitting region. Formally, the line image is resolved with de-convolved size of $250\pm40\times250\pm140$~mas ($1.6\times1.6$~kpc), which is consistent with the size estimate of the continuum emission. The integrated flux density is to be $9.40\pm0.40$~Jy~km~s$^{-1}$, corresponding to a line luminosity of $5.9\times10^9~\mathrm{L}_\odot$.

The velocity vs.~position map (moment~1 map) of the [C\,{\sc II}] line is shown in Panel~C of Fig.~\ref{fig:maps}. We have restricted this analysis to pixels in the data cube where the S/N is at least $4\times$ the rms noise level of $\sim2$~mJy (determined in a line-free channel). The velocity map shows a clear gradient of higher to lower velocity from the east to the west. In the moment~0 map, the S/N is $\sim10$ at the edges of this region, indicating that this gradient is real. This velocity gradient is suggestive of large-scale, ordered rotation; implications are discussed in Section~\ref{sec:discussion}.

In Panel~D of Fig.~\ref{fig:maps}, we present the velocity dispersion map (moment~2 map) of the [C\,{\sc II}] line. The central emission region has a velocity dispersion of about 220~km~s$^{-1}$. On the edges of the emission region, the velocity dispersion drops to $<100$~km~s$^{-1}$. In Section~\ref{subsec:Msigma}, we discuss the meaning of this velocity dispersion according to whether it originates from the host galaxy bulge or from a rotating star-formation disc.

\begin{figure}
\includegraphics[width=3.4in]{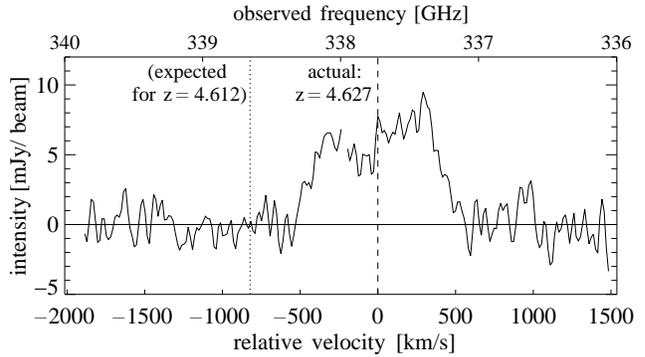}
\caption{ 
ALMA detection of [C\,{\sc II}] line in J1554+1937. The rms variation (determined from line-free channels) is approximately 2~mJy. Observed frequencies are indicated on the top axis. The lower axis shows the velocity relative to the line centre based on a redshift of $z=4.627$. The dashed vertical line indicates the line centre; the dotted vertical line indicates the expected line centre for a redshift of $z=4.612$ (the visually confirmed SDSS redshift). Missing data at 338~GHz correspond to the low sensitivity edges of the spectral windows.
\label{fig:profile}}
\end{figure}

The [C\,{\sc II}] line profile is shown in Fig.~\ref{fig:profile}. The expected redshift was $z=4.612$ (based on the optical spectrum). The actual redshift of the line is approximately 4.627: the observed line centre is 337.8~GHz (887.6~$\mu$m). Because of this offset from the expected frequency, the line falls over the edges of two of ALMA's spectral windows; approximately 19.75~MHz (6.7~km~s$^{-1}$) of data are missing (flagged) at 338~GHz. The profile is clearly non-Gaussian. It is closer to being flat-topped but also shows line structure; it may be a double-horned profile. The FWHM is 735~km~s$^{-1}$.

\section{Analysis}
\label{sec:discussion}

\subsection{Redshift offset of the C\,{\sc IV} and [C\,{\sc II}] lines}
\label{subsec:offset}

We find a redshift offset between the [C\,{\sc II}] emission line and the UV lines in the SDSS spectrum. The [C\,{\sc II}] line is at $z=4.627$ (based on the central frequency of 337.8~GHz), while C\,{\sc IV}, NV, and Ly~$\alpha$ appear to be at lower redshift. The SDSS pipeline \citep{bolton12} measured the redshift of Ly~$\alpha$ to be $z=4.622$; however, its true redshift is difficult to pinpoint owing to asymmetric absorption by the Ly~$\alpha$ forest. We consider the `true' redshift of the ultraviolet lines to be the visually determined redshift reported in SDSS Data Release~9 \citep{dr9}. The redshift determined by eye is $z=4.612$, corresponding to a velocity offset of $\sim800$~km~s$^{-1}$ from [C\,{\sc II}]. Such offsets are not unusual in QSOs. Indeed, \citet{sulentic00} and \citet{richardsEtal02} have observed that the C\,{\sc IV} line is nearly always blue-shifted with respect to low-ionisation lines, with offsets that occasionally exceed 2000~km~s$^{-1}$, although 800~km~s$^{-1}$ is more typical.

One theory put forth to explain the large blueshift in broad, high-ionisation lines like C\,{\sc IV} is commonly referred to as the `disc-wind' model \citep{sulentic00_review,richardsEtal11}. According to this picture, the line is emitted from a radial outflow whose far side is obscured by the accretion disc, and thus only the blue-shifted component is visible. The wind may arise from the disc itself.

We wish to emphasize this caveat to future ALMA observers: in the sub-millimetre regime, such velocity offsets can amount to shifts of $\sim1$~GHz in the observed frame, potentially moving a targeted sub-millimetre line halfway across one of ALMA's 2-GHz-wide spectral windows. If the redshift of a QSO is known only from the (rest-frame) UV lines, optical emission lines, we therefore recommend that ALMA observers place multiple spectral windows alongside each other to allow for an unknown, but potentially quite large, shift in the line frequency.

\subsection{Implications from the [C\,{\sc II}] velocity map} 
\label{subsec:disc}

The velocity map of the [C\,{\sc II}] line (Panel~C of Fig.~\ref{fig:maps}) shows a gradient across the source, with the eastern side of the source being redshifted with respect to the western side. This morphology suggests large-scale rotation of a gravitationally bound gas component. Similar line morphologies were observed in five $z\ga6$ QSOs by \citet{wang13} and \citet{willott13}. In those cases, as for the source presented here, the observations were not high-spatial-resolution, such that the detailed velocity morphology is not tightly constrained. However, the overall gradient is clear.

One possible origin for such a velocity gradient is a rotating gas disc that is observed partially edge-on. Is the velocity dispersion map (Panel~D of Fig.~\ref{fig:maps}) consistent with a rotating disc? Generally, a rotation-supported disc should show a velocity dispersion that is fairly constant across the disc, representing the intrinsic dispersion of the disc material. For an inclined disc, however, poorly resolved rotation can contaminate the dispersion map, an effect known as `beam-smearing' \citep{DaviesEtal11,NewmanEtal13}. A velocity-dispersion peak near the centroid of an unresolved source strengthens the evidence for rotating gas. This morphology is in fact what we observe for J1554+1937 (Panel~D of Fig.~\ref{fig:maps}), which has a velocity dispersion peak of $\sim220$~km~s$^{-1}$ near the source centre. The true intrinsic dispersion is better measured near the outskirts of the emission region, which in this case suggest an intrinsic dispersion of 50--100~km~s$^{-1}$ if the emission is truly from a disc. This level of dispersion is similar to values measured for molecular gas discs by \citet{swinbank11} and \citet{hodge12}. However, there are problems with such an interpretation. Studies of high-redshift galaxy morphology with the {\it Hubble Space Telescope} argue against the presence of well ordered galaxy discs at redshifts above $\sim2$ \citep[e.g.,][]{glazebrook95,driver98,ferguson04,papovich05}. At high redshift, star formation seems to have occurred in massive clumps ($10^7$--$10^9~M_\odot$) within turbulent discs \citep{elmegreenElmegreen05,elmegreen09}. Theoretical formulations \citep[e.g.,][]{dekel09} support the notion that stable discs did not form in galaxies before $z\sim1$.

If the [C\,{\sc II}] emission does not originate in a rotating disc, it may instead be coming from the host galaxy's bulge, or a progenitor to the bulges that are seen in local galaxies \citep[e.g.,][]{elmegreen09}, which would also explain the small spatial extent of high-redshift [C\,{\sc II}] emission. Galaxy bulges in the local Universe are typically `red and dead' (i.e., with little or no gas or star-formation) \citep[e.g.,][]{renzini06review,blanton09review,fabian12review}, but at higher redshift they must have undergone intense star formation in order to generate the massive stellar bulges observed today. The luminous [C\,{\sc II}] line in this and other sources may be due to vigorous star formation in the young bulges, or bulge progenitors, of early-stage galaxies.

Bi-directional outflows of ionized gas are another possible origin for the [C\,{\sc II}] velocity gradient. Outflows have been claimed to be ubiquitous in luminous QSOs, as evidenced by mapping of extended ionized-gas regions \citep[e.g.,][]{liuZakamskaGreene14,mcelroy15} and analysis of broad absorption lines \citep[e.g.,][]{shen08}. In fact, \citet{ganguly07} observed a positive correlation between the presence of broad absorption lines and a QSO's luminosity, suggesting that outflows are more prevalent in the most luminous QSOs. Furthermore, \citet{maiolino12} and \citet{cicone15} observed a striking [C\,{\sc II}] line profile in a $z=6.4$ QSO (one of the most distant known, with bolometric luminosity $8\times10^{12}~\mathrm{L_\odot}$), with broad wings that are attributed to a massive outflow. While the narrow component of the [C\,{\sc II}] line in that source has FWHM of $\sim350$~km~s$^{-1}$, the low-level broad wings are consistent with a Gaussian profile extending to $\pm1300$~km~s$^{-1}$, and are marginally resolved on the physical scale of the host galaxy ($\sim16$~kpc).
If the [C\,{\sc II}] emission in J1554+1937 originates from outflowing material, that material must have a large range of velocities in order to produce the observed line profile shown in Fig.~\ref{fig:profile}. However, as the profile and velocity gradient can also be explained by the presence of a massive rotating disc, invoking outflows as a possible explanation is not strictly necessary. Perhaps the [C\,{\sc II}] emission is actually generated by a combination of disc rotation and outflowing gas.

Another physical process that is associated with luminous QSOs is that of major mergers between gas-rich galaxies. While QSO activity can be induced by secular (internal) processes within a galaxy, major mergers seem to be the driving force behind the most luminous QSOs \citep{sandersMirabel96,urrutiaLacyBecker08,treisterEtal12}. If a merger was the precursor to this extremely luminous QSO, then the gas kinematics likely show signs of disruption or turbulence, which could be revealed by higher-spatial-resolution spectroscopy. Activity resulting from a merger does not remove the possibility that outflowing gas is present. On the contrary, the evolutionary framework describing QSO formation purports that major galaxy mergers trigger a `feedback' phase where energy from the SMBH disrupts the host galaxy's star formation activity \citep{sandersEtal88, kauffmannHaehnelt00, wyitheLoeb03, dimatteoNature05, hopkins08quasars}; outflows are one possible form of QSO feedback.

The obvious next step toward identifying the true nature of the [C\,{\sc II}] line is to obtain higher-spatial-resolution spectroscopy with ALMA. Such observations would allow the isolation of the line profile in different physical regions of the galaxy, a technique known as `kinemetry' \citep[e.g.][]{krajnovic06}. Asymmetries in such mappings have been used to distinguish between disc rotation and merger activity \citep{shapiro08}, while ionisation diagnostics from kinemetry can trace directional outflows \citep{mcelroy15}.

\subsection{[C\,{\sc II}] line profile}
\label{subsec:profile}

Based on the velocity map, we have speculated that the line originates in a massive rotating disc. In this section, we show that a rotating disc can generate a line profile similar to the one shown in Fig.~\ref{fig:profile}. We avoid complex modelling at present, as the low spatial resolution of the current observations does not constrain the possible dynamical theories.

In general, the line profile from a rotating disc depends on the velocity field (i.e., the rotation curve), the surface brightness profile (i.e., the gas distribution), and the inclination of the disc. The surface brightness profile and the inclination determine the line intensity at each position on the projected disc image, while the rotation curve determines the observed velocity of the emission at each position.

Observed emission lines from discs show a huge range of profiles, as demonstrated by neutral hydrogen (HI) \citep{koribalski04}. The two extreme types of line profile are the singly peaked (typically Gaussian) profile of irregular dwarf galaxies and the double-peaked or `double-horned' profiles of large spiral galaxies. Both types of galaxy have exponential surface brightness profiles, but their velocity fields are different. The rotation curve of a typical spiral galaxy initially increases with radius, then levels off at some limiting velocity \citep[e.g.,][]{rubin78,clemens85}. Such galaxies have a double-horned HI profile due to a large range of radii generating emission in a small range of velocities, resulting in symmetric red-shifted and blue-shifted peaks corresponding to the maximum rotational velocity offset from the systemic velocity. In contrast, late-type dwarf galaxies typically have a continually rising velocity curve \citep[e.g.,][]{cote00}, resulting in a single-peaked line at the systemic velocity. In between those two extremes, it is possible to have a nearly flat-topped profile. This profile manifests when the characteristic radius (the `knee') of the surface brightness profile is the same as the characteristic radius of the velocity rotation curve. In other words, the surface brightness falls roughly as the velocity curve rises, resulting in constant line luminosity across a range of observed velocities.

We demonstrate idealized line profiles using a disc-galaxy model with standard rotation curve and surface brightness distribution, as follows. We assume a fiducial velocity ($v$) rotation curve that initially rises exponentially, then levels off with some characteristic radius $r_0$ such that $v(r) = v_0[1 - \exp(-r/r_\mathrm{0})]$. For the surface brightness ($I$), we use an exponential profile with characteristic radius $r_\mathrm{c}$ such that $I(r)\propto\exp(-r/r_\mathrm{c})$, as is seen in normal disc galaxies. For perfectly ordered rotation, the emission line from any point on the disc could be approximated as a delta function. However, the addition of an intrinsic velocity dispersion ($\sigma_\mathrm{v}$) yields a line profile ($L$) at any point on the disc given by 
\begin{equation}
L(\mathrm{v})=\frac{1}{\sqrt{2\pi\sigma_\mathrm{v}^2}}~e^{-\frac{1}{2}\left(\frac{\mathrm{v}-v(r)}{\sigma_\mathrm{v}}\right)^2}, 
\end{equation}
where $v(r)$ corresponds to the average velocity at that radius.
Finally, we allow for an inclination of the disc to the line of sight. The spatially integrated line profile from the disc is therefore the convolution of the line profile and the brightness profile integrated over the (projected) face of the disc.

\begin{figure}
\includegraphics[width=3.3in]{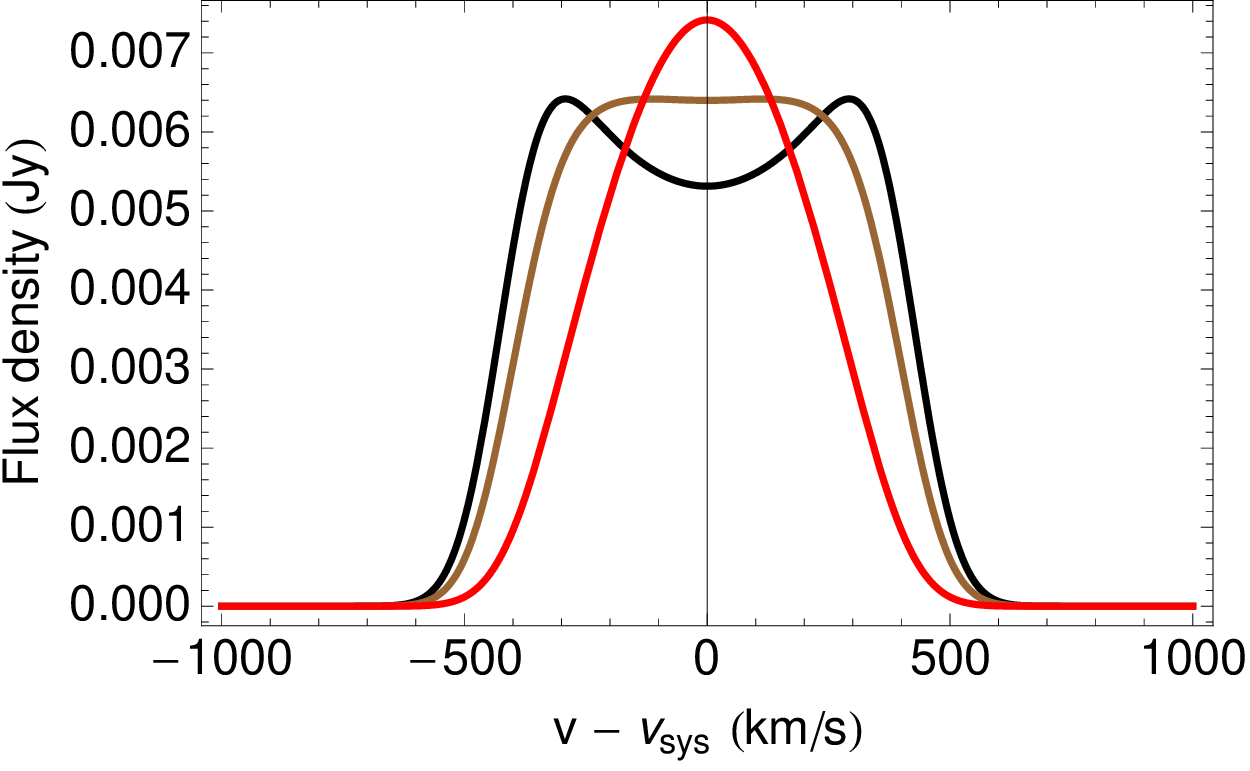}
\caption{Examples of three different line profiles from dynamically equivalent discs. The formalisation of the model is given in the text. For all three profiles here, $v_0=480$~km~s$^{-1}$, $\sigma_v=75$~km~s$^{-1}$, and $i=60\degr$. The value $r_\mathrm{c}/r_0=0.5$ generates the single-peaked (red) profile; $r_\mathrm{c}/r_0=1$ generates the flat-topped (brown) profile; $r_\mathrm{c}/r_0=2$ generates the double-peaked (black) profile. The line amplitudes are arbitrary for the purposes of this figure.
\label{fig:profiles}}
\end{figure}

Example line profiles are shown in Fig.~\ref{fig:profiles}. For the purposes of this demonstration, we assign the parameters $v_0=480$~km~s$^{-1}$, $\sigma_v=75$~km~s$^{-1}$, and $i=60\degr$. A ratio of $r_\mathrm{c}/r_0=0.5$ generates the single-peaked profile, $r_\mathrm{c}/r_0=1$ generates the flat-topped profile, and $r_\mathrm{c}/r_0=2$ generates the double-peaked profile. These three profiles thus correspond to discs with equivalent dynamics; the variation is due to the ratio of $r_\mathrm{c}/r_0$. In other words, the observed profile depends on the surface brightness at the outer edge of the disc, where the rotational velocity is at its maximum.

\begin{figure}
\includegraphics[width=3.3in]{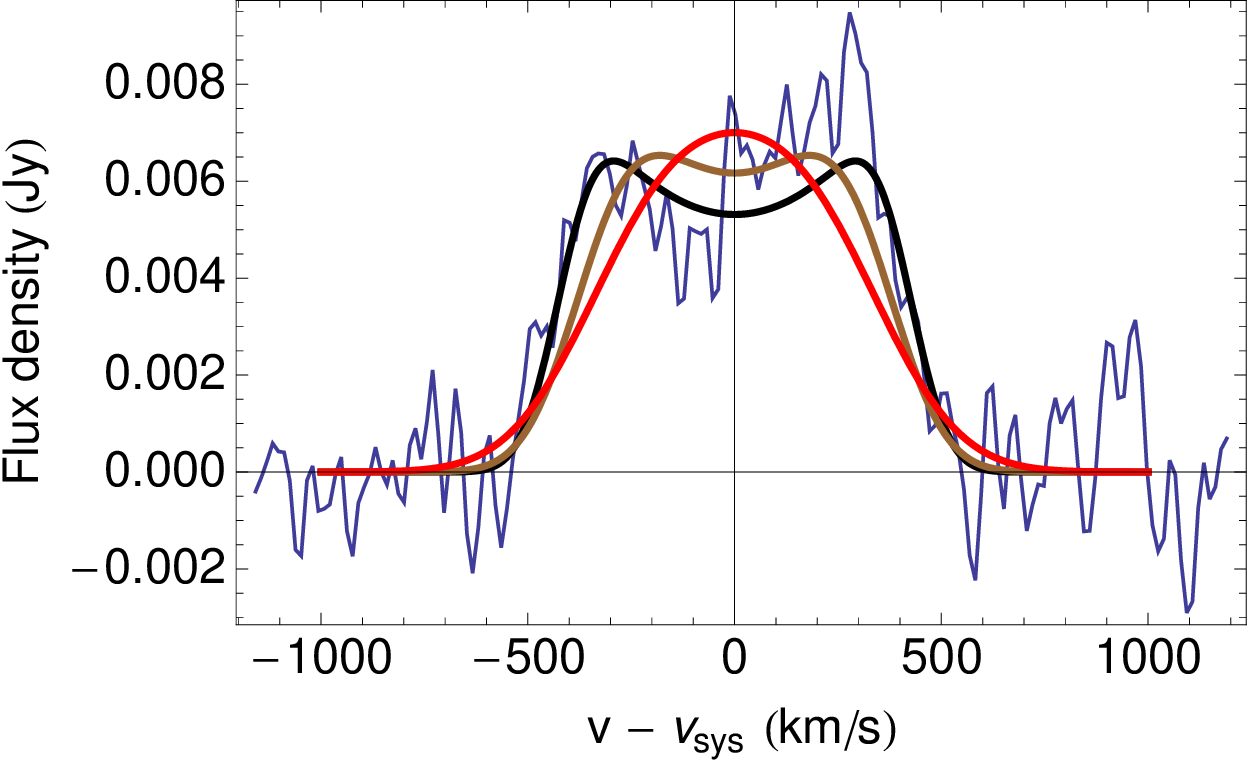}
\caption{Line profiles shown here demonstrate that a large intrinsic velocity dispersion in the disc can mask the double-horned rotation signature, resulting in a single-peaked profile. The formalisation of the model is given in the text. For the three profiles here, $r_\mathrm{c}/r_0=2$ and $i=60\degr$. The maximum velocity ($v_0$) and intrinsic velocity dispersion ($\sigma_v$) in kilometres per second are: $v_0=350$, $\sigma_v=175$ for the single-peaked (red) profile; $v_0=415$, $\sigma_v=110$ for the nearly flat-topped (brown) profile; and $v_0=480$, $\sigma_v=75$ for the strongly double-peaked (black) profile. The purple data points correspond to the [C\,{\sc II}] line profile of J1554+1937 shown in Fig.~\ref{fig:profile}.
\label{fig:fits}}
\end{figure}

A ratio of $r_\mathrm{c}/r_0>1$ is required but not sufficient to generate a double-peaked line profile. A single-peaked profile can result from an intrinsic velocity dispersion that is high enough to mask the double-horned rotation signature. This effect is demonstrated in Fig.~\ref{fig:fits}, which shows three different models corresponding to varying values of maximum velocity ($v_0$) and intrinsic velocity dispersion ($\sigma_v$). Relatively higher intrinsic velocity can dominate over the wings of the line associated with rotation, yielding a Gaussian profile. The fact that the [C\,{\sc II}] line observed in J1554+1937 appears to be double-peaked or flat-topped rules out the notion that the line width is merely a result of high velocity dispersion. 

It is important to emphasize that the parameters for inclination ($i$) and maximum velocity ($v_0$) are degenerate in this model. Therefore, although the model formulated above separates these two parameters, we do not claim to know the true parameters as a result of these rough fits. Qualitatively, the profile shape implies that, if the line is generated by a disc, then the disc has a relatively high maximum velocity and a relatively low intrinsic velocity dispersion.

\subsection{Estimating dynamical mass for a rotating disc}

If we assume a rotating disc geometry for the origin of the [C\,{\sc II}] line, we can estimate the dynamical mass within the disc radius as $M_\mathrm{dyn}/M_\odot\approx1.16\times10^{5}v_0^2D$ where $v_0$ is the maximum circular velocity of the disc in km~s$^{-1}$ and $D$ is the disc diameter in kpc. We visually estimate a maximum circular velocity using the half-width of the line at 20 per cent of its maximum value: $v_0\times\sin(i)\approx480\pm50~$km~s$^{-1}$, where $i$ is the inclination angle of the disc relative to the line of sight ($i=0\degr$ for a face-on disc). Using the source size from the line data fitting ($D\sim1.6$~kpc; see Section~\ref{sec:data}), the estimated dynamical mass is $M_\mathrm{dyn}\times\sin^2(i)\approx(4.3\pm1.2)\times10^{10}~M_\odot$. Using the size estimate from the continuum data ($D\sim2.0$~kpc) yields a slightly larger mass estimate of $M_\mathrm{dyn}\times\sin^2(i)\approx(5.3\pm1.3)\times10^{10}~M_\odot$.
The poor spatial resolution prevents us from determining the inclination, and shape-based inclination estimates are controversial in any case due to unknown intrinsic ellipticity and variations of ellipticity with radius \citep{bellovary14}. Certainly, unless the disc is very close to fully face-on (e.g., $i<20\degr$), the dynamical mass must be in the range of $5\times10^{10}$--$6\times10^{11}~M_\odot$. Thus, this source probably has a dynamical mass similar to the $z\sim6$ QSOs studied by \citet{wang13}; such values are also typical of QSOs at lower redshifts \citep{solomonVandenBoutReview05}. 

We can compare the estimated dynamical mass to that same parameter observed in other gas-rich galaxies, both locally and at high redshift. Rotation curves and dynamical masses are often determined using hydrogen lines such as H$\alpha$, which probes the star-forming regions of a galaxy. For example, the Sydney--AAO Multi-object Integral field spectrograph (SAMI) Galaxy Survey \citep{sami_edr} has produced high-angular-resolution data cubes for a large number of local galaxies with a range of galaxy types. In about 120 galaxies from this sample with $z<0.06$, G.~Cecil et al.~(in prep.) finds maximum dynamical masses of about 10$^{11}~\mathrm{M}_\odot$, with a typical value of about $2\times10^{10}~\mathrm{M}_\odot$. At $z\sim2$, \citet{sinfoni06} observed H$\alpha$ in several UV-selected galaxies and found that they were generally consistent with having clumpy, rotating gas discs within the half-light radius of $\sim4$~kpc, with dynamical masses ranging from $\sim0.5$ to $25\times10^{10}~\mathrm{M}_\odot$. Thus, the mass we estimate for J1554+1937 places it in the high-mass tail of the galaxy population.

\subsection{Estimating central velocity dispersion for a bulge\label{subsec:sigma}}

If the [C\,{\sc II}] line originates from the host galaxy bulge, then we can use the velocity dispersion of the line to investigate the source with respect to the $M$--$\sigma$ relation: the empirical scaling relation that relates the mass ($M_\mathrm{BH}$) of a galaxy's SMBH to the stellar velocity dispersion ($\sigma$) of its bulge component. We first estimate the bulge velocity dispersion in this section; the results are discussed in context of the $M$--$\sigma$ relation in Section~\ref{subsec:Msigma}.

To estimate the stellar velocity dispersion, the simplest approach is to use the velocity dispersion of the [C\,{\sc II}] line directly, a method that was also used by \citet{venemans12} and \citet{willott13} for high-redshift QSOs detected in [C\,{\sc II}]. For J1554+1937, the velocity dispersion map suggests that $\sigma\approx220$~km~s$^{-1}$ (see Panel~D of Fig.~\ref{fig:maps}). We caution that the value we measure may not be the intrinsic bulge velocity dispersion if there are any non-bulge components contributing to the measured line dispersion. For example, \citet{bellovary14} show how disc rotation and inclination effects combine to falsely inflate dispersion measurements. Additionally, any rotating component that is not spatially resolved can increase the velocity dispersion measurement due to `beam-smearing' \citep{DaviesEtal11,NewmanEtal13}. These effects are extremely difficult to account for without knowing the true galaxy inclination.

Does the dispersion of the [C\,{\sc II}] line provide a reliable measurement of the bulge stellar velocity dispersion? For most of the other distant sources with estimated stellar velocity dispersion, that value was determined from the widths of narrow UV/optical emission lines \citep[e.g.,][]{shields03,salviander07}. The [OIII] forbidden line, for example, originates from the narrow-line region surrounding the black hole, and its width correlates well with the stellar velocity dispersion in the bulge \citep{nelson00,boroson03}. That correlation suggests that the velocity field of the narrow-line region is determined in large part by the same gravitational dynamics that govern the stellar velocity field \citep{nelsonWhittle96}. The question of whether [C\,{\sc II}] similarly probes the stellar velocity dispersion depends on the nature of the [C\,{\sc II}]-emitting gas. As discussed above, it is likely that a young galaxy bulge at $z=4.6$ will be experiencing significant star-formation. The [C\,{\sc II}] luminosity is correlated with SFR in normal galaxies \citep[e.g.,][]{hollenbachTielens97,deLooze11}, but in active galaxies its luminosity is thought to be `contaminated' by gas that has been ionized by radiation from the active nucleus \citep[e.g.,][]{malhotraEtal97,ds4mnras}. The [C\,{\sc II}] line need not sample the bulge at all if the bulge region is highly ionized. 

We now return to the possibility that [C\,{\sc II}] originates from a rotating star-forming disc instead of from a bulge region, as discussed in Sections~\ref{subsec:disc} and \ref{subsec:profile}. If the [C\,{\sc II}] line traces a rotating disc, then the velocity dispersion map is likely a result of `beam-smearing'. Even so, it may be possible to obtain a velocity dispersion estimate for the bulge component using the following relation. As discussed in \citet{nelsonWhittle96}, the velocity dispersion of the bulge measures its gravitational potential while the maximum disc rotation velocity measures the potential of the disc and/or halo. Thus, simple theory based on the mass density profiles of galaxies predicts that the ratio between the maximum circular velocity of a galaxy's disc and the velocity dispersion of its bulge should fall in the range $2^{1/2}$--$3^{1/2}$, with the exact value depending on the mass profile.

\citet{ferrarese02} was the first to empirically quantify the relationship between circular rotation velocity ($v_0$) and central velocity dispersion ($\sigma$). Using a nearby sample of 33 elliptical and spiral galaxies, Ferrarese derived a power-law fit in the region where $70<\sigma<180$~km~s$^{-1}$, but noted that for elliptical galaxies it may also be valid in the range $120<\sigma<350$~km~s$^{-1}$. Following that result, \citet{pizzella05} examined a broader sample of 72 elliptical and spiral galaxies to investigate the relation between $\sigma$ and $v_0$ separately for high- and low-surface-brightness galaxies. They found that these two classes of galaxies follow different relations, although the latter was determined from a sample of just eight. In addition, Pizzella et al.~found that the high-surface-brightness and elliptical galaxies follow a single relation that is better characterised by a linear fit in the $v_0$--$\sigma$ plane than by the power-law fit of \citet{ferrarese02}. Their linear relation fits their data well for velocity dispersion as low as $\sigma\sim50$~km~s$^{-1}$, and is a much better fit to the data with $300<\sigma<350$~km~s$^{-1}$ (cf.~their Figure~2) than the relation from \citet{ferrarese02}. We therefore use the linear relation from \citet{pizzella05} to represent the relationship between $\sigma$ and $v_0$. We reproduce their linear fit here as
\begin{equation}
\label{eq:vcir}
\left(\frac{v_0}{\mathrm{km~s}^{-1}}\right) = (1.32\pm0.09)\left(\frac{\sigma_{e/8}}{\mathrm{km~s}^{-1}}\right)+(46\pm14),
\end{equation}
where $v_0$ is determined where the rotation velocity curve is flat, and $\sigma_{e/8}$ signifies that the value of $\sigma$ corresponds to an aperture of size $r_e/8$. The radius $r_e$ is the ``effective radius" of the bulge --- the radius that contains half of the total amount of light \citep{baggett98}.

Without knowing the actual inclination of the disc, we cannot determine the true de-projected maximum rotation velocity, $v_0$. However, we can estimate the maximum {\it projected} circular velocity from the width of the [C\,{\sc II}] line. A reasonable estimate for the projected maximum rotational velocity is the half-width of the line at 20 per cent of its maximum, yielding $v_0=480$~km~s$^{-1}/\sin(i)$; a profile corresponding to this value is shown in Fig.~\ref{fig:fits} as the double-horned black line. Applying this value in Equation~\ref{eq:vcir} yields the approximation that $\sigma_{e/8}\approx330$~km~s$^{-1}/\sin(i)$. However, the {\it projected} maximum rotation velocity is a lower limit to the {\it true} maximum rotation velocity, such that $v_0>480$~km~s$^{-1}$; similarly it follows that $\sigma_{e/8}\ga330$~km~s$^{-1}$. 

In Section~\ref{subsec:Msigma}, we compare the properties of J1554+1937 to the local $M$--$\sigma$ relation as presented by  \citet{kormendyHo_review}. Those authors use a value of $\sigma$ corresponding to an aperture of size $r_e/2$, and therefore we must perform an aperture correction to our $\sigma_{e/8}$ estimate. We apply the correction given by \citet{jorgensenFranxKjaergaard95}, who used kinematic models based on empirical data to derive a formula for the measured central velocity dispersion as a function of aperture size. Using their formulation, we find that $\sigma_{e/2} = 0.93\times\sigma_{e/8}$; therefore $\sigma_{e/2}\ga310$~km~s$^{-1}$ for J1554+1937. For inclination $i=60\degr$, we find $\sigma_{e/2}\approx360$~km~s$^{-1}$, and for $i=45\degr$, we find $\sigma_{e/2}\approx450$~km~s$^{-1}$.

The largest known galaxy bulge velocity dispersions are $\la400$~km~s$^{-1}$ \citep[see references within][]{kormendyHo_review}, which means that J1554+1937 source is in the high-velocity-dispersion tail of the known distribution.
A virialized bulge with $\sigma\sim400$~km~s$^{-1}$ must have a high mass or a small radius, or both. For a simple estimate of the properties of such a bulge, we use the virial theorem, which states: 
\begin{equation}
\label{eq:virial}
M_\mathrm{bulge}\sim\frac{3R\sigma^2}{G},
\end{equation}
where $M_\mathrm{bulge}$ is the bulge mass and $R$ is the virial radius. For this system, the bulge mass must be less than the total dynamical mass, which was calculated to be  $M_\mathrm{dyn}\times\sin^2(i)\approx(5.3\pm1.3)\times10^{10}~M_\odot$. If the bulge has virialized, then Equation~\ref{eq:virial} leads to the conclusion that $R\la0.8$~kpc. Such a bulge is within the typical population distribution for galaxies in the redshift range $1<z<3$ \citet{bruce14a}, although toward the high-mass/low-radius side of that distribution. Such compact cores are rare in nearby galaxies, but it has been demonstrated that the cores of galaxies at high redshift are usually more compact than their local counterparts \citep{vandokkum14}.

From \citet{kormendyHo_review}, we note an important caveat to this analysis. In contrast to the \citet{ferrarese02} results, Kormendy \& Ho assert that the tight correlation between $v_0$ and $\sigma$ only exists in the range of $150$~km~s$^{-1}<v_0<300$~km~s$^{-1}$. At higher rotational velocities, data are scarce and the scatter visibly increases. Therefore, our estimate for bulge velocity dispersion must be taken cautiously. Furthermore, the correlation holds for galaxies with classical bulges (having the light profile of an elliptical galaxy, resulting from galaxy mergers) but not galaxies with so-called ``pseudobulges" (resulting from secular processes). Applying the $v_0/\sigma$ relation to J1554+1937 thus requires the assumption that its host galaxy contains a classical bulge.

\subsection{Measuring the central black hole mass}
\label{subsec:mbh}

\citet{shenEtal11} estimated the SMBH in J1554+1937 to have a mass of $M_\mathrm{BH} = (2.8\pm0.5)\times10^{10}~M_\odot$. This mass estimate was based on the the width of the C\,{\sc IV} line according to the calibration presented by \citet{VP06}. The error published by Shen et al.~represents only the measurement error from the line fitting procedure, which equates to an uncertainty of 0.08~dex. However, Vestergaard \& Peterson assert that the uncertainty in their mass calibrations based on C\,{\sc IV} width is actually 0.32~dex. Based on the latter estimate for uncertainty, we consider the probable mass range for the SMBH in J1554+1937 to be (1.3--5.9)$\times10^{10}~M_\odot$.

\citet{parkEtal13} presented an updated mass calibration for C\,{\sc IV} that would decrease the SMBH mass estimate to $M_\mathrm{BH}=6.3\times10^{9}~M_\odot$. However, \citet{kratzer15} showed that this new calibration is likely to be systematically low by 0.5~dex or more, a conclusion that supports the original mass estimate for this source. While such a mass is toward the high end of what is theoretically observable \citep{kellyEtal10}, it is not unprecedented, as other black holes approximately this massive are known at high redshift \citep[based on other C\,{\sc IV} widths;][]{shenEtal11} as well as locally \citep[based on small-scale stellar velocity dispersions;][]{mcconnell11Nature}.

Does the C\,{\sc IV} line width provide a reliable measurement of the black hole mass? This line is thought to be the least reliable estimator \citep[e.g.,][]{hoEtal12}, but at high redshifts it is usually the only estimator available. For nearby ($\la100$~Mpc) systems, the SMBH mass can be reliably determined by dynamical measurements, whereas higher-redshift characterisations are based on single-epoch measurements of broad-line widths (e.g., H$\alpha$, MgII, or C\,{\sc IV}) that were first calibrated against the local $M$--$\sigma$ relation. Owing to a selection bias at high redshift, single-epoch estimators may falsely conflate SMBH mass estimates \citep{shenKelly10}. Additionally, the C\,{\sc IV} line width might be influenced by outflows, particularly in sources with large offsets between the C\,{\sc IV} redshift and the systemic redshift (see Sec.~\ref{subsec:offset}), or by a major galaxy merger. On the other hand, \citet{finecroom10} have argued that single-epoch estimators actually depend more on intrinsic QSO luminosity than on the line widths themselves. In this case, despite the inherent uncertainty in using C\,{\sc IV} as a mass estimator, the high bolometric luminosity of J1554+1937 indicates that its SMBH must be extremely massive. If the SMBH is Eddington-limited (assuming the classical optically thin case), then its mass is actually much lower: $\sim4.0\times10^{9}~M_\odot$. However, such a luminous system could also result from an even lower-mass SMBH that is radiating at a super-Eddington rate, though such super-Eddington QSOs are expected to be quite rare \citep{kellyEtal10}. If the current mass estimate is correct, then the Eddington ratio ($L_\mathrm{bol}/L_\mathrm{Edd}$) is 0.15. 

\subsection{Comparison to the $M$--$\sigma$ relation}
\label{subsec:Msigma}

\begin{figure}
\includegraphics[width=3.5in]{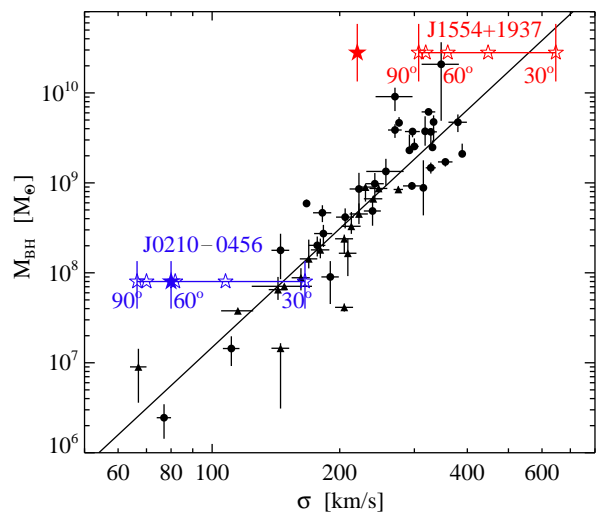}
\caption{
Reproduction of the local $M$--$\sigma$ relation from \citet{kormendyHo_review} (for $\sigma=\sigma_{e/2}$). Filled circles represent local elliptical galaxies and filled triangles represent local spiral galaxies with classical bulges. The filled star symbols represent the locations of J1554+1937 (this paper) and J0210$-$0456 \citep{willott13} if the value of $\sigma$ were equal to the observed the velocity dispersion of the [C\,{\sc II}] emission line. Hollow star symbols indicate our estimates of velocity dispersion (Sec~\ref{subsec:sigma}) for various inclination angles in increments of $15\degr$ (as labelled).
\label{fig:msigma}}
\end{figure}

With estimates in hand for the bulge velocity dispersion (Sec.~\ref{subsec:sigma}) and central black hole mass (Sec~\ref{subsec:mbh}), we are now equipped to compare J1554+1937 to the local $M$--$\sigma$ relation \citep{kormendyHo_review}, which we reproduce here as
\begin{multline}
\label{eq:Msigma}
\log\left(\frac{M_\mathrm{BH}}{M_\odot}\right) = \\ (8.49\pm0.05)+ (4.38\pm0.29)\log\left(\frac{\sigma_{e/2}}{200~\mathrm{km~s^{-1}}}\right).
\end{multline}
In Figure~\ref{fig:msigma}, we show the data and fit presented by Kormendy \& Ho. We show the estimated locations for J1554+1937 for different potential values of the unknown disc inclination, using the black hole mass estimate of $M_\mathrm{BH}=(2.8^{+3.1}_{-1.5})\times10^{10}M_\odot$. For the minimum estimated value of $\sigma_{e/2}=310$~km~s$^{-1}$ (corresponding to $i=90\degr$), we see that J1554+1937 would be over an order of magnitude more massive in $M_\mathrm{BH}$ than predicted by the local relation. Given that the intrinsic scatter of the $M$--$\sigma$ relation is 0.29~dex \citep{kormendyHo_review}, such an offset corresponds to a discrepancy at the level of 2.6$\sigma_\mathrm{rms}$. For inclinations in the range $27\degr<i<45\degr$, J1554+1937 would consistent with the local $M$--$\sigma$ relation within $1\sigma_\mathrm{rms}$. Therefore, for likely inclinations between 45\degr\ and 90\degr, J1554+1937 can be considered over-massive with respect to the relation by an order of magnitude or less, but not at a  significant level.

For comparison, we also show in Figure~\ref{fig:msigma} the location of J1554+1937 if one were to equate the velocity dispersion of the [C\,{\sc II}] line directly with the velocity dispersion of the bulge. In taking that approach, we would find that J1554+1937 is nearly two orders of magnitude above the local $M$--$\sigma$ relation (with a significance of $4\sigma_\mathrm{rms}$).  
That technique is often used for high-redshift QSOs, and is appropriate if the line originates in the bulge. However, for disc-dominated line emission such as that seen in this source, that approach and the resulting conclusion would be incorrect. This is a crucial point, given that  [C\,{\sc II}] emission in other high-redshift sources also shows evidence of large-scale rotation \citep{wang13,willott13,deBreuck14}.

What are the results if we apply the same analysis to other QSOs with [C\,{\sc II}] detections? Unfortunately, there are almost no sources in the literature with equivalent measurements for performing this analysis. Most of the QSOs with a [C\,{\sc II}] detection do not have virial SMBH mass estimates \citep{pety04,maiolino09,gallerani12,wagg12,wang13} or do not show a [C\,{\sc II}] velocity gradient \citep{maiolino05,maiolino12,walter09,venemans12,cicone15}. The one source that allows a direct comparison is CFHQS J0210$-$0456 \citep{willott13}, which has a velocity gradient in the [C\,{\sc II}] line emission as well as a SMBH mass determined from the width of Mg\,II\,$\lambda$2799; like J1554+1937, its inclination is unknown. The SMBH mass is $M=8.0^{+5.5}_{-4.0}\times10^7~M_\odot$ \citep{willott10}. Using a velocity dispersion of $\sigma\sim80$~km~s$^{-1}$ (as estimated from the FWHM--$\sigma$ relationship for a Gaussian distribution), \citet{willott13} found the SMBH in this source to be approximately an order of magnitude more massive than predicted from the local $M$--$\sigma$ relation. When we apply our above analysis to this source, we obtain the following results. The [C\,{\sc II}] line of J0210$-$0456 appears to have a half-width at 20 per cent intensity of $\sim140$~km~s$^{-1}$. Equation~\ref{eq:vcir} suggests a bulge velocity dispersion of $\sigma\sim66$~km~s$^{-1}/\sin(i)$. If the [C\,{\sc II}] line in this source originates from a rotating disk that has an inclination of 60\degr\ to the line of sight, we estimate a value for the bulge velocity dispersion that is similar to what Willott et al.~found by assuming that the line originates from the bulge itself. For more highly inclined disks, this source would be offset by an larger factor than found by Willott et al. Thus, when assuming that the [C\,{\sc II}] line in J0210$-$0456 originates in a rotating disc in a galaxy that also contains a bulge, we agree with the conclusion of \citet{willott13} that the SMBH in this source is likely to be over-massive compared to the local $M$--$\sigma$ relation.

Recently, \citet{willott15} updated their discussion of the $M$--$\sigma$ relation for J0210$-$0456 (and the other known $z>6$ QSOs) using a similar method to the one presented here, by estimating the bulge velocity dispersion from the disc circular rotation velocity. Their approach and ours are based on the same method, but use different formulations; they have applied the formulation of the $v_0$--$\sigma$ correlation presented by \citet{ho07}. They assign an inclination of $i=64\degr$ based on the minor/major axis ratio, and find $\sigma=98\pm20$~km/s; for the same inclination, our method yields an estimate of $\sigma_{e/2}=78$~km/s. Willott et al.~extended their analysis to include other $z>6$ QSOs in the literature, some of which have black hole mass estimates based on the Eddington limit. Where inclination estimates were not  available, they assumed $i=55\degr$, which is appropriate for a population of uniformly/randomly oriented discs. Similar to the results presented here, Willott et al.~conclude that many $z>6$ QSOs may be consistent with the local $M$--$\sigma$ relation, while others have black holes that are much more massive than predicted from the local relation. 

Variations in the $M$--$\sigma$ relation, and particularly its redshift evolution, remain controversial \citep[e.g.,][and references therein]{kormendyBender11}. For example, there are significant complexities related to inconsistencies in velocity dispersion determination \citep{bellovary14}, questionable SMBH mass estimates \citep[][]{finecroom10}, and selection bias at high redshift \citep{lauer07}. The analysis of J1554+1937 presented in this paper depends on empirical relations between bulge and disc properties, which necessitates the assumption that this source has a central bulge typical of other galaxies. However, the extreme luminosity of this source demonstrates that it is atypical, and therefore such relationships might not apply. For example, such a luminous source may be experiencing or recovering from a major merger, and therefore would not behave like a stable system.

If our assumptions are valid, then we have identified an extremely luminous, high-redshift QSO that is likely to be consistent with the local $M$--$\sigma$ relation. This result does not contradict the conclusions of many earlier observations that high-redshift QSOs are generally more massive than the local relation, which have been interpreted as indicating that SMBHs in the early Universe grew more quickly than their host galaxies \citep{peng06,wang10,willott13}. Rather, this source provides an additional data point to probe the extent of this relation's variation, and suggests that some galaxies, like their SMBHs, could have evolved quickly at high redshifts. However, before stronger claims can be made regarding these conclusions, detailed follow-up of this source at higher angular resolution would be valuable, and other extreme sources should be similarly analysed.

\section{Summary}
\label{sec:summary}

We have presented an ALMA detection of an unusual [C\,{\sc II}] line in QSO J1554+1937, one of the most luminous QSOs in the observable Universe, with bolometric luminosity of $L_\mathrm{bol}=1.3\times10^{14}~\mathrm{L_\odot}$. The line presented here has a very broad (FWHM~$\sim735$~km~s$^{-1}$) non-Gaussian profile, and represents the first [C\,{\sc II}] detection of this kind. The central velocity of the line indicates a redshift of $z=4.627$, corresponding to an offset of 800~km~s$^{-1}$ from the redshift of broad UV lines, which is typical for QSOs. The redshift of the [C\,{\sc II}] line seems to be the systemic redshift of the system. The velocity map shows a gradient from the east-northeast to the west-southwest across the 2~kpc spatial extent of the source. However, as the source is barely resolved, it is not currently possible to determine whether the velocity gradient originates from a rotating disc, from bi-directional outflows, or from another origin altogether. The profile of the line is consistent with the hypothesis of a rotating disc origin, with a maximum circular velocity of $\sim480$~km~s$^{-1}$.

Following the hypothesis that the [C\,{\sc II}] line originates from a rotating disc, we employed an empirical relation to estimate the bulge velocity dispersion in the QSO host galaxy, which requires the implicit assumption that the host galaxy contains a classical central bulge. The results of this analysis suggest that this source is likely consistent with the local $M$--$\sigma$ relation, or is offset from the relation by no more than one order of magnitude in black hole mass. This result stands in contrast to earlier results for other high-redshift QSO sources, whose black holes tend to be one-to-two orders of magnitude more massive than predicted by this relation.Our results are consistent with those of \citet{willott15}, who used a similar approach to update estimates of velocity dispersion for $z>6$ QSOs.

Our result stresses the importance of taking into account the true dynamical origin of the [C\,{\sc II}] line when using it to estimate host galaxy parameters. If we were to use its velocity dispersion as a proxy for the host galaxy velocity dispersion (equivalent to assuming the line originates from a host galaxy bulge), we would have reached a very different conclusion about whether this source is offset from the local $M$--$\sigma$ relation. This analysis is important because numerous high-redshift sources have [C\,{\sc II}] profiles that are suggestive of large-scale rotation of the emitting region.s

A thorough understanding of the [C\,{\sc II}] line in J1554+1937 requires higher-spatial-resolution observations, which are possible using an extended configuration of ALMA. ALMA has only recently opened this new window into the high-redshift Universe, where high-angular-resolution observations of sub-millimetre lines can quickly probe galaxy kinematics at high redshift. Similar observations of other sources will likely follow, enabling us to determine whether this type of emission line is truly rare, or if the line presented here is simply the first of many to follow.

\section*{acknowledgements}

The authors extend thanks to Alister Graham, Ron Ekers, B\"arbel Koribalski, and Scott Croom for thoughtful discussions on this topic. We are grateful to the anonymous reviewer, whose careful reading and comments helped to improve the paper.

This paper makes use of the following ALMA data: ADS/JAO.ALMA\#2012.1.00604.S. ALMA is a partnership of ESO (representing its member states), NSF (USA) and NINS (Japan), together with NRC (Canada), NSC and ASIAA (Taiwan), and KASI (Republic of Korea), in cooperation with the Republic of Chile. The Joint ALMA Observatory is operated by ESO, AUI/NRAO and NAOJ.

The National Radio Astronomy Observatory is a facility of the National Science Foundation operated under cooperative agreement by Associated Universities, Inc.  Funding for the SDSS and SDSS-II has been provided by the Alfred P. Sloan Foundation, the Participating Institutions, the National Science Foundation, the U.S. Department of Energy, the National Aeronautics and Space Administration, the Japanese Monbukagakusho, the Max Planck Society, and the Higher Education Funding Council for England. The SDSS Web Site is http://www.sdss.org/.

Funding for SDSS-III has been provided by the Alfred P. Sloan Foundation, the Participating Institutions, the National Science Foundation, and the U.S. Department of Energy Office of Science. The SDSS-III web site is http://www.sdss3.org/.

SDSS-III is managed by the Astrophysical Research Consortium for the Participating Institutions of the SDSS-III Collaboration including the University of Arizona, the Brazilian Participation Group, Brookhaven National Laboratory, University of Cambridge, Carnegie Mellon University, University of Florida, the French Participation Group, the German Participation Group, Harvard University, the Instituto de Astrofisica de Canarias, the Michigan State/Notre Dame/JINA Participation Group, Johns Hopkins University, Lawrence Berkeley National Laboratory, Max Planck Institute for Astrophysics, Max Planck Institute for Extraterrestrial Physics, New Mexico State University, New York University, Ohio State University, Pennsylvania State University, University of Portsmouth, Princeton University, the Spanish Participation Group, University of Tokyo, University of Utah, Vanderbilt University, University of Virginia, University of Washington, and Yale University.

This publication makes use of data products from the Wide-field Infrared Survey Explorer, which is a joint project of the University of California, Los Angeles, and the Jet Propulsion Laboratory/California Institute of Technology, and NEOWISE, which is a project of the Jet Propulsion Laboratory/California Institute of Technology. WISE and NEOWISE are funded by the National Aeronautics and Space Administration.

This research has made use of the NASA/IPAC Extragalactic Database (NED), which is operated by the Jet Propulsion Laboratory, California Institute of Technology, under contract with the National Aeronautics and Space Administration.
\bibliography{bibliography}

\begin{thebibliography}{}

\bibitem[\protect\citeauthoryear{{Ahn}, {Alexandroff}, {Allende Prieto},
  {Anderson}, {Anderton}, {Andrews}, {Aubourg}, {Bailey}, {Balbinot}, {Barnes}
  \& et al.}{{Ahn} et~al.}{2012}]{dr9}
{Ahn} C.~P.,  {Alexandroff} R.,  {Allende Prieto} C.,  {Anderson} S.~F.,
  {Anderton} T.,  {Andrews} B.~H.,  {Aubourg} {\'E}.,  {Bailey} S.,  {Balbinot}
  E.,  {Barnes} R.,    et al. 2012, ApJS, 203, 21

\bibitem[\protect\citeauthoryear{{Allen}, {Croom}, {Konstantopoulos}, {Bryant},
  {Sharp}, {Cecil}, {Fogarty}, {Foster}, {Green}, {Ho} \& et al.}{{Allen}
  et~al.}{2015}]{sami_edr}
{Allen} J.~T.,  {Croom} S.~M.,  {Konstantopoulos} I.~S.,  {Bryant} J.~J.,
  {Sharp} R.,  {Cecil} G.~N.,  {Fogarty} L.~M.~R.,  {Foster} C.,  {Green}
  A.~W.,  {Ho} I.-T.,    et al. 2015, \mnras, 446, 1567

\bibitem[\protect\citeauthoryear{{Baggett}, {Baggett} \& {Anderson}}{{Baggett}
  et~al.}{1998}]{baggett98}
{Baggett} W.~E.,  {Baggett} S.~M.,    {Anderson} K.~S.~J.,  1998, \aj, 116,
  1626

\bibitem[\protect\citeauthoryear{{Becker}, {White} \& {Helfand}}{{Becker}
  et~al.}{1995}]{first}
{Becker} R.~H.,  {White} R.~L.,    {Helfand} D.~J.,  1995, \apj, 450, 559

\bibitem[\protect\citeauthoryear{{Bellovary}, {Holley-Bockelmann},
  {G{\"u}ltekin}, {Christensen}, {Governato}, {Brooks}, {Loebman} \&
  {Munshi}}{{Bellovary} et~al.}{2014}]{bellovary14}
{Bellovary} J.~M.,  {Holley-Bockelmann} K.,  {G{\"u}ltekin} K.,  {Christensen}
  C.~R.,  {Governato} F.,  {Brooks} A.~M.,  {Loebman} S.,    {Munshi} F.,
  2014, \mnras, 445, 2667

\bibitem[\protect\citeauthoryear{{Blanton} \& {Moustakas}}{{Blanton} \&
  {Moustakas}}{2009}]{blanton09review}
{Blanton} M.~R.,  {Moustakas} J.,  2009, \araa, 47, 159

\bibitem[\protect\citeauthoryear{{Bolton}, {Schlegel}, {Aubourg}, {Bailey},
  {Bhardwaj}, {Brownstein}, {Burles}, {Chen}, {Dawson}, {Eisenstein} \& et
  al.}{{Bolton} et~al.}{2012}]{bolton12}
{Bolton} A.~S.,  {Schlegel} D.~J.,  {Aubourg} {\'E}.,  {Bailey} S.,  {Bhardwaj}
  V.,  {Brownstein} J.~R.,  {Burles} S.,  {Chen} Y.-M.,  {Dawson} K.,
  {Eisenstein} D.~J.,    et al. 2012, \aj, 144, 144

\bibitem[\protect\citeauthoryear{{Boroson}}{{Boroson}}{2003}]{boroson03}
{Boroson} T.~A.,  2003, \apj, 585, 647

\bibitem[\protect\citeauthoryear{{Bridge}, {Blain}, {Borys}, {Petty},
  {Benford}, {Eisenhardt}, {Farrah}, {Griffith}, {Jarrett}, {Lonsdale},
  {Stanford}, {Stern}, {Tsai}, {Wright} \& {Wu}}{{Bridge}
  et~al.}{2013}]{bridge13}
{Bridge} C.~R.,  {Blain} A.,  {Borys} C.~J.~K.,  {Petty} S.,  {Benford} D.,
  {Eisenhardt} P.,  {Farrah} D.,  {Griffith} R.~L.,  {Jarrett} T.,  {Lonsdale}
  C.,  {Stanford} S.~A.,  {Stern} D.,  {Tsai} C.-W.,  {Wright} E.~L.,    {Wu}
  J.,  2013, \apj, 769, 91

\bibitem[\protect\citeauthoryear{{Bruce}, {Dunlop}, {McLure}, {Cirasuolo},
  {Buitrago}, {Bowler}, {Targett}, {Bell}, {McIntosh}, {Dekel}, {Faber},
  {Ferguson}, {Grogin}, {Hartley}, {Kocevski}, {Koekemoer}, {Koo} \&
  {McGrath}}{{Bruce} et~al.}{2014}]{bruce14a}
{Bruce} V.~A.,  {Dunlop} J.~S.,  {McLure} R.~J.,  {Cirasuolo} M.,  {Buitrago}
  F.,  {Bowler} R.~A.~A.,  {Targett} T.~A.,  {Bell} E.~F.,  {McIntosh} D.~H.,
  {Dekel} A.,  {Faber} S.~M.,  {Ferguson} H.~C.,  {Grogin} N.~A.,  {Hartley}
  W.,  {Kocevski} D.~D.,  {Koekemoer} A.~M.,  {Koo} D.~C.,    {McGrath} E.~J.,
  2014, \mnras, 444, 1001

\bibitem[\protect\citeauthoryear{{Carilli} \& {Walter}}{{Carilli} \&
  {Walter}}{2013}]{carilli_review13}
{Carilli} C.~L.,  {Walter} F.,  2013, \araa, 51, 105

\bibitem[\protect\citeauthoryear{{Cicone}, {Maiolino}, {Gallerani}, {Neri},
  {Ferrara}, {Sturm}, {Fiore}, {Piconcelli} \& {Feruglio}}{{Cicone}
  et~al.}{2015}]{cicone15}
{Cicone} C.,  {Maiolino} R.,  {Gallerani} S.,  {Neri} R.,  {Ferrara} A.,
  {Sturm} E.,  {Fiore} F.,  {Piconcelli} E.,    {Feruglio} C.,  2015, \aap,
  574, A14

\bibitem[\protect\citeauthoryear{{Clemens}}{{Clemens}}{1985}]{clemens85}
{Clemens} D.~P.,  1985, \apj, 295, 422

\bibitem[\protect\citeauthoryear{{Condon}, {Cotton}, {Greisen}, {Yin},
  {Perley}, {Taylor} \& {Broderick}}{{Condon} et~al.}{1998}]{nvss}
{Condon} J.~J.,  {Cotton} W.~D.,  {Greisen} E.~W.,  {Yin} Q.~F.,  {Perley}
  R.~A.,  {Taylor} G.~B.,    {Broderick} J.~J.,  1998, \aj, 115, 1693

\bibitem[\protect\citeauthoryear{{C{\^o}t{\'e}}, {Carignan} \&
  {Freeman}}{{C{\^o}t{\'e}} et~al.}{2000}]{cote00}
{C{\^o}t{\'e}} S.,  {Carignan} C.,    {Freeman} K.~C.,  2000, \aj, 120, 3027

\bibitem[\protect\citeauthoryear{{Davies}, {F{\"o}rster Schreiber}, {Cresci},
  {Genzel}, {Bouch{\'e}}, {Burkert}, {Buschkamp}, {Genel}, {Hicks}, {Kurk},
  {Lutz}, {Newman}, {Shapiro}, {Sternberg}, {Tacconi} \& {Wuyts}}{{Davies}
  et~al.}{2011}]{DaviesEtal11}
{Davies} R.,  {F{\"o}rster Schreiber} N.~M.,  {Cresci} G.,  {Genzel} R.,
  {Bouch{\'e}} N.,  {Burkert} A.,  {Buschkamp} P.,  {Genel} S.,  {Hicks} E.,
  {Kurk} J.,  {Lutz} D.,  {Newman} S.,  {Shapiro} K.,  {Sternberg} A.,
  {Tacconi} L.~J.,    {Wuyts} S.,  2011, \apj, 741, 69

\bibitem[\protect\citeauthoryear{{De Breuck}, {Williams}, {Swinbank},
  {Caselli}, {Coppin}, {Davis}, {Maiolino}, {Nagao}, {Smail}, {Walter},
  {Wei{\ss}} \& {Zwaan}}{{De Breuck} et~al.}{2014}]{deBreuck14}
{De Breuck} C.,  {Williams} R.~J.,  {Swinbank} M.,  {Caselli} P.,  {Coppin} K.,
   {Davis} T.~A.,  {Maiolino} R.,  {Nagao} T.,  {Smail} I.,  {Walter} F.,
  {Wei{\ss}} A.,    {Zwaan} M.~A.,  2014, \aap, 565, A59

\bibitem[\protect\citeauthoryear{{de Looze}, {Baes}, {Bendo}, {Cortese} \&
  {Fritz}}{{de Looze} et~al.}{2011}]{deLooze11}
{de Looze} I.,  {Baes} M.,  {Bendo} G.~J.,  {Cortese} L.,    {Fritz} J.,  2011,
  \mnras, 416, 2712

\bibitem[\protect\citeauthoryear{{Dekel}, {Sari} \& {Ceverino}}{{Dekel}
  et~al.}{2009}]{dekel09}
{Dekel} A.,  {Sari} R.,    {Ceverino} D.,  2009, \apj, 703, 785

\bibitem[\protect\citeauthoryear{{Di Matteo}, {Springel} \& {Hernquist}}{{Di
  Matteo} et~al.}{2005}]{dimatteoNature05}
{Di Matteo} T.,  {Springel} V.,    {Hernquist} L.,  2005, \nat, 433, 604

\bibitem[\protect\citeauthoryear{{D{\'{\i}}az-Santos}, {Armus}, {Charmandaris},
  {Stierwalt}, {Murphy}, {Haan}, {Inami}, {Malhotra}, {Meijerink}, {Stacey},
  {Petric}, {Evans}, {Veilleux}, {van der Werf}, {Lord} \& et
  al.}{{D{\'{\i}}az-Santos} et~al.}{2013}]{ds4mnras}
{D{\'{\i}}az-Santos} T.,  {Armus} L.,  {Charmandaris} V.,  {Stierwalt} S.,
  {Murphy} E.~J.,  {Haan} S.,  {Inami} H.,  {Malhotra} S.,  {Meijerink} R.,
  {Stacey} G.,  {Petric} A.~O.,  {Evans} A.~S.,  {Veilleux} S.,  {van der Werf}
  P.~P.,  {Lord} S.,    et al. 2013, \apj, 774, 68

\bibitem[\protect\citeauthoryear{{Driver}, {Fern{\'a}ndez-Soto}, {Couch},
  {Odewahn}, {Windhorst}, {Phillipps}, {Lanzetta} \& {Yahil}}{{Driver}
  et~al.}{1998}]{driver98}
{Driver} S.~P.,  {Fern{\'a}ndez-Soto} A.,  {Couch} W.~J.,  {Odewahn} S.~C.,
  {Windhorst} R.~A.,  {Phillipps} S.,  {Lanzetta} K.,    {Yahil} A.,  1998,
  \apjl, 496, L93

\bibitem[\protect\citeauthoryear{{Eisenhardt}, {Wu}, {Tsai}, {Assef},
  {Benford}, {Blain}, {Bridge}, {Condon}, {Cushing}, {Cutri}, {Evans} II,
  {Gelino}, {Griffith}, {Grillmair}, {Jarrett}, {Lonsdale}, {Masci}, {Mason} \&
  {Petty}}{{Eisenhardt} et~al.}{2012}]{eis4mnras}
{Eisenhardt} P.~R.~M.,  {Wu} J.,  {Tsai} C.-W.,  {Assef} R.,  {Benford} D.,
  {Blain} A.,  {Bridge} C.,  {Condon} J.~J.,  {Cushing} M.~C.,  {Cutri} R.,
  {Evans} II N.~J.,  {Gelino} C.,  {Griffith} R.~L.,  {Grillmair} C.~J.,
  {Jarrett} T.,  {Lonsdale} C.~J.,  {Masci} F.~J.,  {Mason} B.~S.,    {Petty}
  S. e.~a.,  2012, \apj, 755, 173

\bibitem[\protect\citeauthoryear{{Elmegreen} \& {Elmegreen}}{{Elmegreen} \&
  {Elmegreen}}{2005}]{elmegreenElmegreen05}
{Elmegreen} B.~G.,  {Elmegreen} D.~M.,  2005, \apj, 627, 632

\bibitem[\protect\citeauthoryear{{Elmegreen}, {Elmegreen}, {Fernandez} \&
  {Lemonias}}{{Elmegreen} et~al.}{2009}]{elmegreen09}
{Elmegreen} B.~G.,  {Elmegreen} D.~M.,  {Fernandez} M.~X.,    {Lemonias} J.~J.,
   2009, \apj, 692, 12

\bibitem[\protect\citeauthoryear{{Fabian}}{{Fabian}}{2012}]{fabian12review}
{Fabian} A.~C.,  2012, \araa, 50, 455

\bibitem[\protect\citeauthoryear{{Ferguson}, {Dickinson}, {Giavalisco},
  {Kretchmer}, {Ravindranath}, {Idzi}, {Taylor}, {Conselice}, {Fall},
  {Gardner}, {Livio}, {Madau}, {Moustakas}, {Papovich}, {Somerville}, {Spinrad}
  \& {Stern}}{{Ferguson} et~al.}{2004}]{ferguson04}
{Ferguson} H.~C.,  {Dickinson} M.,  {Giavalisco} M.,  {Kretchmer} C.,
  {Ravindranath} S.,  {Idzi} R.,  {Taylor} E.,  {Conselice} C.~J.,  {Fall}
  S.~M.,  {Gardner} J.~P.,  {Livio} M.,  {Madau} P.,  {Moustakas} L.~A.,
  {Papovich} C.~M.,  {Somerville} R.~S.,  {Spinrad} H.,    {Stern} D.,  2004,
  \apjl, 600, L107

\bibitem[\protect\citeauthoryear{{Ferrarese}}{{Ferrarese}}{2002}]{ferrarese02}
{Ferrarese} L.,  2002, \apj, 578, 90

\bibitem[\protect\citeauthoryear{{Ferrarese} \& {Merritt}}{{Ferrarese} \&
  {Merritt}}{2000}]{ferrareseMerritt00}
{Ferrarese} L.,  {Merritt} D.,  2000, \apjl, 539, L9

\bibitem[\protect\citeauthoryear{{Fine}, {Croom}, {Bland-Hawthorn}, {Pimbblet},
  {Ross}, {Schneider} \& {Shanks}}{{Fine} et~al.}{2010}]{finecroom10}
{Fine} S.,  {Croom} S.~M.,  {Bland-Hawthorn} J.,  {Pimbblet} K.~A.,  {Ross}
  N.~P.,  {Schneider} D.~P.,    {Shanks} T.,  2010, \mnras, 409, 591

\bibitem[\protect\citeauthoryear{{F{\"o}rster Schreiber}, {Genzel}, {Lehnert},
  {Bouch{\'e}}, {Verma}, {Erb}, {Shapley}, {Steidel}, {Davies}, {Lutz} \& et
  al.}{{F{\"o}rster Schreiber} et~al.}{2006}]{sinfoni06}
{F{\"o}rster Schreiber} N.~M.,  {Genzel} R.,  {Lehnert} M.~D.,  {Bouch{\'e}}
  N.,  {Verma} A.,  {Erb} D.~K.,  {Shapley} A.~E.,  {Steidel} C.~C.,  {Davies}
  R.,  {Lutz} D.,    et al. 2006, \apj, 645, 1062

\bibitem[\protect\citeauthoryear{{Gallerani}, {Neri}, {Maiolino},
  {Mart{\'{\i}}n}, {De Breuck}, {Walter}, {Caselli}, {Krips}, {Meneghetti},
  {Nagao}, {Wagg} \& {Walmsley}}{{Gallerani} et~al.}{2012}]{gallerani12}
{Gallerani} S.,  {Neri} R.,  {Maiolino} R.,  {Mart{\'{\i}}n} S.,  {De Breuck}
  C.,  {Walter} F.,  {Caselli} P.,  {Krips} M.,  {Meneghetti} M.,  {Nagao} T.,
  {Wagg} J.,    {Walmsley} M.,  2012, \aap, 543, A114

\bibitem[\protect\citeauthoryear{{Ganguly}, {Brotherton}, {Cales}, {Scoggins},
  {Shang} \& {Vestergaard}}{{Ganguly} et~al.}{2007}]{ganguly07}
{Ganguly} R.,  {Brotherton} M.~S.,  {Cales} S.,  {Scoggins} B.,  {Shang} Z.,
  {Vestergaard} M.,  2007, \apj, 665, 990

\bibitem[\protect\citeauthoryear{{Glazebrook}, {Ellis}, {Santiago} \&
  {Griffiths}}{{Glazebrook} et~al.}{1995}]{glazebrook95}
{Glazebrook} K.,  {Ellis} R.,  {Santiago} B.,    {Griffiths} R.,  1995, \mnras,
  275, L19

\bibitem[\protect\citeauthoryear{{Graham} \& {Li}}{{Graham} \&
  {Li}}{2009}]{graham09}
{Graham} A.~W.,  {Li} I.-h.,  2009, \apj, 698, 812

\bibitem[\protect\citeauthoryear{{Graham}, {Onken}, {Athanassoula} \&
  {Combes}}{{Graham} et~al.}{2011}]{graham11}
{Graham} A.~W.,  {Onken} C.~A.,  {Athanassoula} E.,    {Combes} F.,  2011,
  \mnras, 412, 2211

\bibitem[\protect\citeauthoryear{{Graham} \& {Scott}}{{Graham} \&
  {Scott}}{2015}]{graham15}
{Graham} A.~W.,  {Scott} N.,  2015, \apj, 798, 54

\bibitem[\protect\citeauthoryear{{Green}}{{Green}}{2006}]{green06}
{Green} P.~J.,  2006, \apj, 644, 733

\bibitem[\protect\citeauthoryear{{G{\"u}ltekin}, {Richstone}, {Gebhardt},
  {Lauer}, {Tremaine}, {Aller}, {Bender}, {Dressler}, {Faber}, {Filippenko},
  {Green}, {Ho}, {Kormendy}, {Magorrian}, {Pinkney} \& {Siopis}}{{G{\"u}ltekin}
  et~al.}{2009}]{gultekinRelations09}
{G{\"u}ltekin} K.,  {Richstone} D.~O.,  {Gebhardt} K.,  {Lauer} T.~R.,
  {Tremaine} S.,  {Aller} M.~C.,  {Bender} R.,  {Dressler} A.,  {Faber} S.~M.,
  {Filippenko} A.~V.,  {Green} R.,  {Ho} L.~C.,  {Kormendy} J.,  {Magorrian}
  J.,  {Pinkney} J.,    {Siopis} C.,  2009, \apj, 698, 198

\bibitem[\protect\citeauthoryear{{Hao}, {Strauss}, {Tremonti}, {Schlegel},
  {Heckman}, {Kauffmann}, {Blanton}, {Fan}, {Gunn}, {Hall} \& et al.}{{Hao}
  et~al.}{2005}]{hao05}
{Hao} L.,  {Strauss} M.~A.,  {Tremonti} C.~A.,  {Schlegel} D.~J.,  {Heckman}
  T.~M.,  {Kauffmann} G.,  {Blanton} M.~R.,  {Fan} X.,  {Gunn} J.~E.,  {Hall}
  P.~B.,    et al. 2005, \aj, 129, 1783

\bibitem[\protect\citeauthoryear{{Heckman} \& {Best}}{{Heckman} \&
  {Best}}{2014}]{heckmanBest14}
{Heckman} T.~M.,  {Best} P.~N.,  2014, \araa, 52, 589

\bibitem[\protect\citeauthoryear{{Ho}}{{Ho}}{2007}]{ho07}
{Ho} L.~C.,  2007, \apj, 669, 821

\bibitem[\protect\citeauthoryear{{Ho}, {Goldoni}, {Dong}, {Greene} \&
  {Ponti}}{{Ho} et~al.}{2012}]{hoEtal12}
{Ho} L.~C.,  {Goldoni} P.,  {Dong} X.-B.,  {Greene} J.~E.,    {Ponti} G.,
  2012, \apj, 754, 11

\bibitem[\protect\citeauthoryear{{Hodge}, {Carilli}, {Walter}, {de Blok},
  {Riechers}, {Daddi} \& {Lentati}}{{Hodge} et~al.}{2012}]{hodge12}
{Hodge} J.~A.,  {Carilli} C.~L.,  {Walter} F.,  {de Blok} W.~J.~G.,  {Riechers}
  D.,  {Daddi} E.,    {Lentati} L.,  2012, \apj, 760, 11

\bibitem[\protect\citeauthoryear{{Hollenbach} \& {Tielens}}{{Hollenbach} \&
  {Tielens}}{1997}]{hollenbachTielens97}
{Hollenbach} D.~J.,  {Tielens} A.~G.~G.~M.,  1997, \araa, 35, 179

\bibitem[\protect\citeauthoryear{{Hopkins}, {Hernquist}, {Cox} \& {Kere{\v
  s}}}{{Hopkins} et~al.}{2008}]{hopkins08quasars}
{Hopkins} P.~F.,  {Hernquist} L.,  {Cox} T.~J.,    {Kere{\v s}} D.,  2008,
  \apjs, 175, 356

\bibitem[\protect\citeauthoryear{{Jahnke} \& {Macci{\`o}}}{{Jahnke} \&
  {Macci{\`o}}}{2011}]{jahnkeMaccio11}
{Jahnke} K.,  {Macci{\`o}} A.~V.,  2011, \apj, 734, 92

\bibitem[\protect\citeauthoryear{{Jorgensen}, {Franx} \&
  {Kjaergaard}}{{Jorgensen} et~al.}{1995}]{jorgensenFranxKjaergaard95}
{Jorgensen} I.,  {Franx} M.,    {Kjaergaard} P.,  1995, \mnras, 276, 1341

\bibitem[\protect\citeauthoryear{{Kauffmann} \& {Haehnelt}}{{Kauffmann} \&
  {Haehnelt}}{2000}]{kauffmannHaehnelt00}
{Kauffmann} G.,  {Haehnelt} M.,  2000, \mnras, 311, 576

\bibitem[\protect\citeauthoryear{{Kellermann}, {Sramek}, {Schmidt}, {Shaffer}
  \& {Green}}{{Kellermann} et~al.}{1989}]{kellermann89}
{Kellermann} K.~I.,  {Sramek} R.,  {Schmidt} M.,  {Shaffer} D.~B.,    {Green}
  R.,  1989, \aj, 98, 1195

\bibitem[\protect\citeauthoryear{{Kelly}, {Vestergaard}, {Fan}, {Hopkins},
  {Hernquist} \& {Siemiginowska}}{{Kelly} et~al.}{2010}]{kellyEtal10}
{Kelly} B.~C.,  {Vestergaard} M.,  {Fan} X.,  {Hopkins} P.,  {Hernquist} L.,
  {Siemiginowska} A.,  2010, \apj, 719, 1315

\bibitem[\protect\citeauthoryear{{Kocevski}, {Faber}, {Mozena}, {Koekemoer},
  {Nandra}, {Rangel}, {Laird}, {Brusa}, {Wuyts}, {Trump} \& et al.}{{Kocevski}
  et~al.}{2012}]{CANDELS12}
{Kocevski} D.~D.,  {Faber} S.~M.,  {Mozena} M.,  {Koekemoer} A.~M.,  {Nandra}
  K.,  {Rangel} C.,  {Laird} E.~S.,  {Brusa} M.,  {Wuyts} S.,  {Trump} J.~R.,
   et al. 2012, \apj, 744, 148

\bibitem[\protect\citeauthoryear{{Koribalski}, {Staveley-Smith}, {Kilborn},
  {Ryder}, {Kraan-Korteweg}, {Ryan-Weber}, {Ekers}, {Jerjen}, {Henning},
  {Putman} \& et al.}{{Koribalski} et~al.}{2004}]{koribalski04}
{Koribalski} B.~S.,  {Staveley-Smith} L.,  {Kilborn} V.~A.,  {Ryder} S.~D.,
  {Kraan-Korteweg} R.~C.,  {Ryan-Weber} E.~V.,  {Ekers} R.~D.,  {Jerjen} H.,
  {Henning} P.~A.,  {Putman} M.~E.,    et al. 2004, \aj, 128, 16

\bibitem[\protect\citeauthoryear{{Kormendy} \& {Bender}}{{Kormendy} \&
  {Bender}}{2011}]{kormendyBender11}
{Kormendy} J.,  {Bender} R.,  2011, \nat, 469, 377

\bibitem[\protect\citeauthoryear{{Kormendy} \& {Ho}}{{Kormendy} \&
  {Ho}}{2013}]{kormendyHo_review}
{Kormendy} J.,  {Ho} L.~C.,  2013, \araa, 51, 511

\bibitem[\protect\citeauthoryear{{Kormendy} \& {Richstone}}{{Kormendy} \&
  {Richstone}}{1995}]{kormendyRichstone95}
{Kormendy} J.,  {Richstone} D.,  1995, \araa, 33, 581

\bibitem[\protect\citeauthoryear{{Krajnovi{\'c}}, {Cappellari}, {de Zeeuw} \&
  {Copin}}{{Krajnovi{\'c}} et~al.}{2006}]{krajnovic06}
{Krajnovi{\'c}} D.,  {Cappellari} M.,  {de Zeeuw} P.~T.,    {Copin} Y.,  2006,
  \mnras, 366, 787

\bibitem[\protect\citeauthoryear{{Kratzer} \& {Richards}}{{Kratzer} \&
  {Richards}}{2015}]{kratzer15}
{Kratzer} R.~M.,  {Richards} G.~T.,  2015, \aj, 149, 61

\bibitem[\protect\citeauthoryear{{Krawczyk}, {Richards}, {Mehta}, {Vogeley},
  {Gallagher}, {Leighly}, {Ross} \& {Schneider}}{{Krawczyk}
  et~al.}{2013}]{krawczyk13}
{Krawczyk} C.~M.,  {Richards} G.~T.,  {Mehta} S.~S.,  {Vogeley} M.~S.,
  {Gallagher} S.~C.,  {Leighly} K.~M.,  {Ross} N.~P.,    {Schneider} D.~P.,
  2013, \apjs, 206, 4

\bibitem[\protect\citeauthoryear{{Krongold}, {Binette} \&
  {Hern{\'a}ndez-Ibarra}}{{Krongold} et~al.}{2010}]{krongold10}
{Krongold} Y.,  {Binette} L.,    {Hern{\'a}ndez-Ibarra} F.,  2010, \apjl, 724,
  L203

\bibitem[\protect\citeauthoryear{{Lauer}, {Tremaine}, {Richstone} \&
  {Faber}}{{Lauer} et~al.}{2007}]{lauer07}
{Lauer} T.~R.,  {Tremaine} S.,  {Richstone} D.,    {Faber} S.~M.,  2007, \apj,
  670, 249

\bibitem[\protect\citeauthoryear{{Liu}, {Zakamska} \& {Greene}}{{Liu}
  et~al.}{2014}]{liuZakamskaGreene14}
{Liu} G.,  {Zakamska} N.~L.,    {Greene} J.~E.,  2014, \mnras, 442, 1303

\bibitem[\protect\citeauthoryear{{Luhman}, {Satyapal}, {Fischer}, {Wolfire},
  {Cox}, {Lord}, {Smith}, {Stacey} \& {Unger}}{{Luhman}
  et~al.}{1998}]{luhman98}
{Luhman} M.~L.,  {Satyapal} S.,  {Fischer} J.,  {Wolfire} M.~G.,  {Cox} P.,
  {Lord} S.~D.,  {Smith} H.~A.,  {Stacey} G.~J.,    {Unger} S.~J.,  1998,
  \apjl, 504, L11

\bibitem[\protect\citeauthoryear{{Maiolino}, {Caselli}, {Nagao}, {Walmsley},
  {De Breuck} \& {Meneghetti}}{{Maiolino} et~al.}{2009}]{maiolino09}
{Maiolino} R.,  {Caselli} P.,  {Nagao} T.,  {Walmsley} M.,  {De Breuck} C.,
  {Meneghetti} M.,  2009, \aap, 500, L1

\bibitem[\protect\citeauthoryear{{Maiolino}, {Cox}, {Caselli}, {Beelen},
  {Bertoldi}, {Carilli}, {Kaufman}, {Menten}, {Nagao}, {Omont}, {Wei{\ss}},
  {Walmsley} \& {Walter}}{{Maiolino} et~al.}{2005}]{maiolino05}
{Maiolino} R.,  {Cox} P.,  {Caselli} P.,  {Beelen} A.,  {Bertoldi} F.,
  {Carilli} C.~L.,  {Kaufman} M.~J.,  {Menten} K.~M.,  {Nagao} T.,  {Omont} A.,
   {Wei{\ss}} A.,  {Walmsley} C.~M.,    {Walter} F.,  2005, \aap, 440, L51

\bibitem[\protect\citeauthoryear{{Maiolino}, {Gallerani}, {Neri}, {Cicone},
  {Ferrara}, {Genzel}, {Lutz}, {Sturm}, {Tacconi}, {Walter}, {Feruglio},
  {Fiore} \& {Piconcelli}}{{Maiolino} et~al.}{2012}]{maiolino12}
{Maiolino} R.,  {Gallerani} S.,  {Neri} R.,  {Cicone} C.,  {Ferrara} A.,
  {Genzel} R.,  {Lutz} D.,  {Sturm} E.,  {Tacconi} L.~J.,  {Walter} F.,
  {Feruglio} C.,  {Fiore} F.,    {Piconcelli} E.,  2012, \mnras, 425, L66

\bibitem[\protect\citeauthoryear{{Malhotra}, {Helou}, {Stacey}, {Hollenbach},
  {Lord}, {Beichman}, {Dinerstein}, {Hunter}, {Lo}, {Lu}, {Rubin},
  {Silbermann}, {Thronson} Jr. \& {Werner}}{{Malhotra}
  et~al.}{1997}]{malhotraEtal97}
{Malhotra} S.,  {Helou} G.,  {Stacey} G.,  {Hollenbach} D.,  {Lord} S.,
  {Beichman} C.~A.,  {Dinerstein} H.,  {Hunter} D.~A.,  {Lo} K.~Y.,  {Lu}
  N.~Y.,  {Rubin} R.~H.,  {Silbermann} N.,  {Thronson} Jr. H.~A.,    {Werner}
  M.~W.,  1997, \apjl, 491, L27

\bibitem[\protect\citeauthoryear{{McConnell}, {Ma}, {Gebhardt}, {Wright},
  {Murphy}, {Lauer}, {Graham} \& {Richstone}}{{McConnell}
  et~al.}{2011}]{mcconnell11Nature}
{McConnell} N.~J.,  {Ma} C.-P.,  {Gebhardt} K.,  {Wright} S.~A.,  {Murphy}
  J.~D.,  {Lauer} T.~R.,  {Graham} J.~R.,    {Richstone} D.~O.,  2011, \nat,
  480, 215

\bibitem[\protect\citeauthoryear{{McElroy}, {Croom}, {Pracy}, {Sharp}, {Ho} \&
  {Medling}}{{McElroy} et~al.}{2015}]{mcelroy15}
{McElroy} R.,  {Croom} S.~M.,  {Pracy} M.,  {Sharp} R.,  {Ho} I.-T.,
  {Medling} A.~M.,  2015, \mnras, 446, 2186

\bibitem[\protect\citeauthoryear{{McLure}, {Jarvis}, {Targett}, {Dunlop} \&
  {Best}}{{McLure} et~al.}{2006}]{mclure06}
{McLure} R.~J.,  {Jarvis} M.~J.,  {Targett} T.~A.,  {Dunlop} J.~S.,    {Best}
  P.~N.,  2006, \mnras, 368, 1395

\bibitem[\protect\citeauthoryear{{McMullin}, {Waters}, {Schiebel}, {Young} \&
  {Golap}}{{McMullin} et~al.}{2007}]{casa}
{McMullin} J.~P.,  {Waters} B.,  {Schiebel} D.,  {Young} W.,    {Golap} K.,
  2007, in {Shaw} R.~A.,  {Hill} F.,   {Bell} D.~J.,  eds, Astronomical Data
  Analysis Software and Systems XVI Vol.~376 of Astronomical Society of the
  Pacific Conference Series, {CASA Architecture and Applications}.
p.~127

\bibitem[\protect\citeauthoryear{{Nelson}}{{Nelson}}{2000}]{nelson00}
{Nelson} C.~H.,  2000, \apjl, 544, L91

\bibitem[\protect\citeauthoryear{{Nelson} \& {Whittle}}{{Nelson} \&
  {Whittle}}{1996}]{nelsonWhittle96}
{Nelson} C.~H.,  {Whittle} M.,  1996, \apj, 465, 96

\bibitem[\protect\citeauthoryear{{Newman}, {Genzel}, {F{\"o}rster Schreiber},
  {Shapiro Griffin}, {Mancini}, {Lilly}, {Renzini}, {Bouch{\'e}}, {Burkert},
  {Buschkamp} \& et al.}{{Newman} et~al.}{2013}]{NewmanEtal13}
{Newman} S.~F.,  {Genzel} R.,  {F{\"o}rster Schreiber} N.~M.,  {Shapiro
  Griffin} K.,  {Mancini} C.,  {Lilly} S.~J.,  {Renzini} A.,  {Bouch{\'e}} N.,
  {Burkert} A.,  {Buschkamp} P.,    et al. 2013, \apj, 767, 104

\bibitem[\protect\citeauthoryear{{Papovich}, {Dickinson}, {Giavalisco},
  {Conselice} \& {Ferguson}}{{Papovich} et~al.}{2005}]{papovich05}
{Papovich} C.,  {Dickinson} M.,  {Giavalisco} M.,  {Conselice} C.~J.,
  {Ferguson} H.~C.,  2005, \apj, 631, 101

\bibitem[\protect\citeauthoryear{{Park}, {Woo}, {Denney} \& {Shin}}{{Park}
  et~al.}{2013}]{parkEtal13}
{Park} D.,  {Woo} J.-H.,  {Denney} K.~D.,    {Shin} J.,  2013, \apj, 770, 87

\bibitem[\protect\citeauthoryear{{Peng}}{{Peng}}{2007}]{peng07}
{Peng} C.~Y.,  2007, \apj, 671, 1098

\bibitem[\protect\citeauthoryear{{Peng}, {Impey}, {Rix}, {Kochanek}, {Keeton},
  {Falco}, {Leh{\'a}r} \& {McLeod}}{{Peng} et~al.}{2006}]{peng06}
{Peng} C.~Y.,  {Impey} C.~D.,  {Rix} H.-W.,  {Kochanek} C.~S.,  {Keeton} C.~R.,
   {Falco} E.~E.,  {Leh{\'a}r} J.,    {McLeod} B.~A.,  2006, \apj, 649, 616

\bibitem[\protect\citeauthoryear{{Pety}, {Beelen}, {Cox}, {Downes}, {Omont},
  {Bertoldi} \& {Carilli}}{{Pety} et~al.}{2004}]{pety04}
{Pety} J.,  {Beelen} A.,  {Cox} P.,  {Downes} D.,  {Omont} A.,  {Bertoldi} F.,
    {Carilli} C.~L.,  2004, \aap, 428, L21

\bibitem[\protect\citeauthoryear{{Pizzella}, {Corsini}, {Dalla Bont{\`a}},
  {Sarzi}, {Coccato} \& {Bertola}}{{Pizzella} et~al.}{2005}]{pizzella05}
{Pizzella} A.,  {Corsini} E.~M.,  {Dalla Bont{\`a}} E.,  {Sarzi} M.,  {Coccato}
  L.,    {Bertola} F.,  2005, \apj, 631, 785

\bibitem[\protect\citeauthoryear{{Reines}, {Sivakoff}, {Johnson} \&
  {Brogan}}{{Reines} et~al.}{2011}]{reines11nature}
{Reines} A.~E.,  {Sivakoff} G.~R.,  {Johnson} K.~E.,    {Brogan} C.~L.,  2011,
  \nat, 470, 66

\bibitem[\protect\citeauthoryear{{Renzini}}{{Renzini}}{2006}]{renzini06review}
{Renzini} A.,  2006, \araa, 44, 141

\bibitem[\protect\citeauthoryear{{Richards}, {Kruczek}, {Gallagher}, {Hall},
  {Hewett}, {Leighly}, {Deo}, {Kratzer} \& {Shen}}{{Richards}
  et~al.}{2011}]{richardsEtal11}
{Richards} G.~T.,  {Kruczek} N.~E.,  {Gallagher} S.~C.,  {Hall} P.~B.,
  {Hewett} P.~C.,  {Leighly} K.~M.,  {Deo} R.~P.,  {Kratzer} R.~M.,    {Shen}
  Y.,  2011, \aj, 141, 167

\bibitem[\protect\citeauthoryear{{Richards}, {Vanden Berk}, {Reichard}, {Hall},
  {Schneider}, {SubbaRao}, {Thakar} \& {York}}{{Richards}
  et~al.}{2002}]{richardsEtal02}
{Richards} G.~T.,  {Vanden Berk} D.~E.,  {Reichard} T.~A.,  {Hall} P.~B.,
  {Schneider} D.~P.,  {SubbaRao} M.,  {Thakar} A.~R.,    {York} D.~G.,  2002,
  \aj, 124, 1

\bibitem[\protect\citeauthoryear{{Rubin}, {Ford} W.~K. \& {Thonnard}}{{Rubin}
  et~al.}{1978}]{rubin78}
{Rubin} V.~C.,  {Ford} W.~K. J.,    {Thonnard} N.,  1978, \apjl, 225, 107

\bibitem[\protect\citeauthoryear{{Salviander}, {Shields}, {Gebhardt} \&
  {Bonning}}{{Salviander} et~al.}{2007}]{salviander07}
{Salviander} S.,  {Shields} G.~A.,  {Gebhardt} K.,    {Bonning} E.~W.,  2007,
  \apj, 662, 131

\bibitem[\protect\citeauthoryear{{Sanders} \& {Mirabel}}{{Sanders} \&
  {Mirabel}}{1996}]{sandersMirabel96}
{Sanders} D.~B.,  {Mirabel} I.~F.,  1996, \araa, 34, 749

\bibitem[\protect\citeauthoryear{{Sanders}, {Soifer}, {Elias}, {Madore},
  {Matthews}, {Neugebauer} \& {Scoville}}{{Sanders}
  et~al.}{1988}]{sandersEtal88}
{Sanders} D.~B.,  {Soifer} B.~T.,  {Elias} J.~H.,  {Madore} B.~F.,  {Matthews}
  K.,  {Neugebauer} G.,    {Scoville} N.~Z.,  1988, \apj, 325, 74

\bibitem[\protect\citeauthoryear{{Schneider}, {Richards}, {Hall}, {Strauss},
  {Anderson}, {Boroson}, {Ross}, {Shen}, {Brandt}, {Fan} \& et al.}{{Schneider}
  et~al.}{2010}]{dr7quasars}
{Schneider} D.~P.,  {Richards} G.~T.,  {Hall} P.~B.,  {Strauss} M.~A.,
  {Anderson} S.~F.,  {Boroson} T.~A.,  {Ross} N.~P.,  {Shen} Y.,  {Brandt}
  W.~N.,  {Fan} X.,    et al. 2010, \aj, 139, 2360

\bibitem[\protect\citeauthoryear{{Shapiro}, {Genzel}, {F{\"o}rster Schreiber},
  {Tacconi}, {Bouch{\'e}}, {Cresci}, {Davies}, {Eisenhauer}, {Johansson},
  {Krajnovi{\'c}} \& et al.}{{Shapiro} et~al.}{2008}]{shapiro08}
{Shapiro} K.~L.,  {Genzel} R.,  {F{\"o}rster Schreiber} N.~M.,  {Tacconi}
  L.~J.,  {Bouch{\'e}} N.,  {Cresci} G.,  {Davies} R.,  {Eisenhauer} F.,
  {Johansson} P.~H.,  {Krajnovi{\'c}} D.,    et al. 2008, \apj, 682, 231

\bibitem[\protect\citeauthoryear{{Shen} \& {Kelly}}{{Shen} \&
  {Kelly}}{2010}]{shenKelly10}
{Shen} Y.,  {Kelly} B.~C.,  2010, \apj, 713, 41

\bibitem[\protect\citeauthoryear{{Shen}, {Richards}, {Strauss}, {Hall},
  {Schneider}, {Snedden}, {Bizyaev}, {Brewington}, {Malanushenko},
  {Malanushenko}, {Oravetz}, {Pan} \& {Simmons}}{{Shen}
  et~al.}{2011}]{shenEtal11}
{Shen} Y.,  {Richards} G.~T.,  {Strauss} M.~A.,  {Hall} P.~B.,  {Schneider}
  D.~P.,  {Snedden} S.,  {Bizyaev} D.,  {Brewington} H.,  {Malanushenko} V.,
  {Malanushenko} E.,  {Oravetz} D.,  {Pan} K.,    {Simmons} A.,  2011, \apjs,
  194, 45

\bibitem[\protect\citeauthoryear{{Shen}, {Strauss}, {Hall}, {Schneider}, {York}
  \& {Bahcall}}{{Shen} et~al.}{2008}]{shen08}
{Shen} Y.,  {Strauss} M.~A.,  {Hall} P.~B.,  {Schneider} D.~P.,  {York} D.~G.,
    {Bahcall} N.~A.,  2008, \apj, 677, 858

\bibitem[\protect\citeauthoryear{{Shields}, {Gebhardt}, {Salviander}, {Wills},
  {Xie}, {Brotherton}, {Yuan} \& {Dietrich}}{{Shields}
  et~al.}{2003}]{shields03}
{Shields} G.~A.,  {Gebhardt} K.,  {Salviander} S.,  {Wills} B.~J.,  {Xie} B.,
  {Brotherton} M.~S.,  {Yuan} J.,    {Dietrich} M.,  2003, \apj, 583, 124

\bibitem[\protect\citeauthoryear{{Solomon} \& {Vanden Bout}}{{Solomon} \&
  {Vanden Bout}}{2005}]{solomonVandenBoutReview05}
{Solomon} P.~M.,  {Vanden Bout} P.~A.,  2005, \araa, 43, 677

\bibitem[\protect\citeauthoryear{{Sulentic}, {Marziani} \&
  {Dultzin-Hacyan}}{{Sulentic} et~al.}{2000}]{sulentic00_review}
{Sulentic} J.~W.,  {Marziani} P.,    {Dultzin-Hacyan} D.,  2000, \araa, 38, 521

\bibitem[\protect\citeauthoryear{{Sulentic}, {Zwitter}, {Marziani} \&
  {Dultzin-Hacyan}}{{Sulentic} et~al.}{2000}]{sulentic00}
{Sulentic} J.~W.,  {Zwitter} T.,  {Marziani} P.,    {Dultzin-Hacyan} D.,  2000,
  \apjl, 536, L5

\bibitem[\protect\citeauthoryear{{Swinbank}, {Papadopoulos}, {Cox}, {Krips},
  {Ivison}, {Smail}, {Thomson}, {Neri}, {Richard} \& {Ebeling}}{{Swinbank}
  et~al.}{2011}]{swinbank11}
{Swinbank} A.~M.,  {Papadopoulos} P.~P.,  {Cox} P.,  {Krips} M.,  {Ivison}
  R.~J.,  {Smail} I.,  {Thomson} A.~P.,  {Neri} R.,  {Richard} J.,    {Ebeling}
  H.,  2011, \apj, 742, 11

\bibitem[\protect\citeauthoryear{{Treister}, {Natarajan}, {Sanders}, {Urry},
  {Schawinski} \& {Kartaltepe}}{{Treister} et~al.}{2010}]{treister10}
{Treister} E.,  {Natarajan} P.,  {Sanders} D.~B.,  {Urry} C.~M.,  {Schawinski}
  K.,    {Kartaltepe} J.,  2010, Science, 328, 600

\bibitem[\protect\citeauthoryear{{Treister}, {Schawinski}, {Urry} \&
  {Simmons}}{{Treister} et~al.}{2012}]{treisterEtal12}
{Treister} E.,  {Schawinski} K.,  {Urry} C.~M.,    {Simmons} B.~D.,  2012,
  \apjl, 758, L39

\bibitem[\protect\citeauthoryear{{Urrutia}, {Lacy} \& {Becker}}{{Urrutia}
  et~al.}{2008}]{urrutiaLacyBecker08}
{Urrutia} T.,  {Lacy} M.,    {Becker} R.~H.,  2008, \apj, 674, 80

\bibitem[\protect\citeauthoryear{{van Dokkum}, {Bezanson}, {van der Wel},
  {Nelson}, {Momcheva}, {Skelton}, {Whitaker}, {Brammer}, {Conroy},
  {F{\"o}rster Schreiber} \& et al.}{{van Dokkum} et~al.}{2014}]{vandokkum14}
{van Dokkum} P.~G.,  {Bezanson} R.,  {van der Wel} A.,  {Nelson} E.~J.,
  {Momcheva} I.,  {Skelton} R.~E.,  {Whitaker} K.~E.,  {Brammer} G.,  {Conroy}
  C.,  {F{\"o}rster Schreiber} N.~M.,    et al. 2014, \apj, 791, 45

\bibitem[\protect\citeauthoryear{{Venemans}, {McMahon}, {Walter}, {Decarli},
  {Cox}, {Neri}, {Hewett}, {Mortlock}, {Simpson} \& {Warren}}{{Venemans}
  et~al.}{2012}]{venemans12}
{Venemans} B.~P.,  {McMahon} R.~G.,  {Walter} F.,  {Decarli} R.,  {Cox} P.,
  {Neri} R.,  {Hewett} P.,  {Mortlock} D.~J.,  {Simpson} C.,    {Warren} S.~J.,
   2012, \apjl, 751, L25

\bibitem[\protect\citeauthoryear{{Vestergaard} \& {Peterson}}{{Vestergaard} \&
  {Peterson}}{2006}]{VP06}
{Vestergaard} M.,  {Peterson} B.~M.,  2006, \apj, 641, 689

\bibitem[\protect\citeauthoryear{{Wagg}, {Wiklind}, {Carilli}, {Espada},
  {Peck}, {Riechers}, {Walter}, {Wootten}, {Aravena}, {Barkats} \& et
  al.}{{Wagg} et~al.}{2012}]{wagg12}
{Wagg} J.,  {Wiklind} T.,  {Carilli} C.~L.,  {Espada} D.,  {Peck} A.,
  {Riechers} D.,  {Walter} F.,  {Wootten} A.,  {Aravena} M.,  {Barkats} D.,
  et al. 2012, \apjl, 752, L30

\bibitem[\protect\citeauthoryear{{Walter}, {Riechers}, {Cox}, {Neri},
  {Carilli}, {Bertoldi}, {Weiss} \& {Maiolino}}{{Walter}
  et~al.}{2009}]{walter09}
{Walter} F.,  {Riechers} D.,  {Cox} P.,  {Neri} R.,  {Carilli} C.,  {Bertoldi}
  F.,  {Weiss} A.,    {Maiolino} R.,  2009, \nat, 457, 699

\bibitem[\protect\citeauthoryear{{Wang}, {Carilli}, {Neri}, {Riechers}, {Wagg},
  {Walter}, {Bertoldi}, {Menten}, {Omont}, {Cox} \& {Fan}}{{Wang}
  et~al.}{2010}]{wang10}
{Wang} R.,  {Carilli} C.~L.,  {Neri} R.,  {Riechers} D.~A.,  {Wagg} J.,
  {Walter} F.,  {Bertoldi} F.,  {Menten} K.~M.,  {Omont} A.,  {Cox} P.,
  {Fan} X.,  2010, \apj, 714, 699

\bibitem[\protect\citeauthoryear{{Wang}, {Wagg}, {Carilli}, {Walter},
  {Lentati}, {Fan}, {Riechers}, {Bertoldi}, {Narayanan}, {Strauss}, {Cox},
  {Omont}, {Menten}, {Knudsen}, {Neri} \& {Jiang}}{{Wang}
  et~al.}{2013}]{wang13}
{Wang} R.,  {Wagg} J.,  {Carilli} C.~L.,  {Walter} F.,  {Lentati} L.,  {Fan}
  X.,  {Riechers} D.~A.,  {Bertoldi} F.,  {Narayanan} D.,  {Strauss} M.~A.,
  {Cox} P.,  {Omont} A.,  {Menten} K.~M.,  {Knudsen} K.~K.,  {Neri} R.,
  {Jiang} L.,  2013, \apj, 773, 44

\bibitem[\protect\citeauthoryear{{Willott}, {Albert}, {Arzoumanian},
  {Bergeron}, {Crampton}, {Delorme}, {Hutchings}, {Omont}, {Reyl{\'e}} \&
  {Schade}}{{Willott} et~al.}{2010}]{willott10}
{Willott} C.~J.,  {Albert} L.,  {Arzoumanian} D.,  {Bergeron} J.,  {Crampton}
  D.,  {Delorme} P.,  {Hutchings} J.~B.,  {Omont} A.,  {Reyl{\'e}} C.,
  {Schade} D.,  2010, \aj, 140, 546

\bibitem[\protect\citeauthoryear{{Willott}, {Bergeron} \& {Omont}}{{Willott}
  et~al.}{2015}]{willott15}
{Willott} C.~J.,  {Bergeron} J.,    {Omont} A.,  2015, \apj, 801, 123

\bibitem[\protect\citeauthoryear{{Willott}, {Omont} \& {Bergeron}}{{Willott}
  et~al.}{2013}]{willott13}
{Willott} C.~J.,  {Omont} A.,    {Bergeron} J.,  2013, \apj, 770, 13

\bibitem[\protect\citeauthoryear{{Wu}, {Tsai}, {Sayers}, {Benford}, {Bridge},
  {Blain}, {Eisenhardt}, {Stern}, {Petty}, {Assef} \& et al.}{{Wu}
  et~al.}{2012}]{wu12_HLIRG}
{Wu} J.,  {Tsai} C.-W.,  {Sayers} J.,  {Benford} D.,  {Bridge} C.,  {Blain} A.,
   {Eisenhardt} P.~R.~M.,  {Stern} D.,  {Petty} S.,  {Assef} R.,    et al.
  2012, \apj, 756, 96

\bibitem[\protect\citeauthoryear{{Wyithe} \& {Loeb}}{{Wyithe} \&
  {Loeb}}{2003}]{wyitheLoeb03}
{Wyithe} J.~S.~B.,  {Loeb} A.,  2003, \apj, 595, 614

\end{thebibliography}

\end{document}